\documentclass[twocolumn]{aastex631}

\usepackage{newtxtext,newtxmath}
\usepackage{graphicx}	
\usepackage{amsmath}	
\usepackage{amssymb}	

\def\gtorder{\mathrel{\raise.3ex\hbox{$>$}\mkern-14mu
    \lower0.6ex\hbox{$\sim$}}}
\def\ltorder{\mathrel{\raise.3ex\hbox{$<$}\mkern-14mu
    \lower0.6ex\hbox{$\sim$}}}

\begin{document}

\title{Direct Collapse Accretion Disks Within Dark Matter Halos:\\ Saturation of the Magnetorotational Instability and the Field Expulsion}
\shorttitle{Direct Collapse Disks: Saturation of the MRI and Field Expulsion}



\author[0000-0002-2243-2790]{Yang~Luo}
\affiliation{Department of Astronomy, Yunnan University, Kunming, Yunnan 650091, China} 

\author[0000-0002-1233-445X]{Isaac~Shlosman}
\affiliation{Department of Physics \& Astronomy, University of Kentucky, Lexington, KY 40506-0055, USA} 
\affiliation{Theoretical Astrophysics, Department of Earth \& Space Science, Osaka University, Osaka 560-0043, Japan} 


\email{luoyang@ynu.edu.cn (YL)}
\email{isaac.shlosman@uky.edu (IS)}

\shortauthors{Luo and Shlosman}

\begin{abstract}
We have used high-resolution zoom-in simulations of direct collapse to supermassive black hole (SMBH) seeds within dark mater (DM) halos in the presence of magnetic fields generated during the collapse, down to $10^{-5}$\,pc or 2\,AU. We confirm an efficient amplification of magnetic field during collapse, the formation of a geometrically thick self-gravitating accretion disk inside 0.1\,pc, and damping of fragmentation in the disk by the field. This disk differs profoundly from SMBH accretion disks. We find the following: (1) The accretion disk is subject to the magnetorotational instability which further amplifies the field to near equipartition. No artificial seeding of the disk field has been used. (2) The equipartition toroidal field changes its polarity in the midplane. (3) The nonlinear Parker instability develops, accompanied by the vertical buckling of the field lines, which injects material above the disk, leading to an increase in the disk scale height; (4) With the Coriolis force producing a coherent helicity above the disk, vertical poloidal field has been generated and amplified. (5) We estimate that the associated outflow will be most probably squashed by  accretion. The resulting configuration consists of a magnetized disk with $\beta \gtorder 0.1$ and its magnetosphere with $\beta << 1$, where $\beta = P_{\rm th}/P_{\rm B}$ is the ratio of thermal to magnetic energy density. (6) The disk is highly variable, due to feeding by variable accretion flow, and strong vortical motions are present. (7) Finally, the negative gradient of the total vertical stress drives an {\it equatorial} outflow sandwiched by an inward accretion flow. 
\end{abstract}




\section{Introduction} 
\label{sec:intro}
Detection of supermassive black holes (SMBHs) of $\sim 10^9\,M_\odot$, only few hundred million years after the Big Bang \citep[e.g., ][]{wu15,venemans17,banados18,bosman23}, does not leave many options for their formation --- direct baryonic collapse within dark matter (DM) halos, Population\,III remnants, or ultra-compact stellar clusters. Extremely fine-tuned conditions or unreasonably long super-Eddington growth rate make the latter two options highly improbable, even after plausible discovery of lower mass SMBHs, $\sim 4\times 10^7\,M_\odot$ at $z\sim 10.3$ \citep{bogdan23}, $\sim 10^7\,M_\odot$ at $z\sim 8.7$ \citep{larson23}, and $\sim 2\times 10^6\,M_\odot$ at $z\sim 10.5$ \citep{maiolino23}. But difficulties exist also in understanding various stages of direct collapse leading to the SMBH seeds. These issues include the SMBH growth at lower redshifts as well \citep[e.g.,][]{ruzmaikin88,pudritz89,begelman23,bhowmick24}. 

The direct collapse scenario allows to circumvent many of the pitfalls but rises new questions that are being gradually addressed \citep[e.g.,][and refs. therein]{inayoshi20}. Formation of SMBHs requires an efficient loss of angular momentum by the accreting gas. Such a process must involve application of either gravitational or magnetic torques or both. The former can develop naturally during gravitational collapse within DM halos and/or as a result of disk instabilities \citep[e.g.,][]{shlosman89,lodato07,begelman09,choi13,choi15}, while the latter can be associated with the developing turbulence which powers the small-scale dynamo during this collapse \citep[e.g.,][]{kazantsev68,kraichnan68,brandenburg96,subramanian98,turk09,federrath11,steinwandel19,begelman23b,latif23b}, or the action of a mean-field $\alpha-\Omega$ dynamo \citep[e.g.,][]{ruzmaikin79}. Since \citet{zeldovich83}, a long list of numerical simulations have confirmed the exponential amplification of the seed field in the largest turbulent scales until its saturation around the equipartition. 

Magnetic torques can be especially important if the central mass accumulation has developed, because in this case the gravitational torques' effect becomes too weak. Therefore, inserting a massive sink particle in a non-MHD collapse in DM halo will lead directly to the formation of a disk around it due to vanishing torques \citep{shlosman16}.  

Magnetized disks have attracted attention already for a long time. Even a weak field can generate turbulence and grow, leading to magnetic viscosity and the magneto-rotational instability (MRI), as originally pointed out by \citet{velihov59} and \citet{chandrasekhar60}, and developed by \citet{balbus91,balbus98}. But importance of magnetic fields goes well beyond the torques. Abundant literature over the last half a century have worked out various effects in hydromagnetic (MHD) fluids. Magnetic fields can be the source of a local viscosity in accretion disks \citep[e.g.,][]{shakura73,novikov73,tan04,brandenburg05,silk06,machida06,sur08,sharda21} and nonlocal viscosity \citep[e.g.,][]{blandford82,emmering92}. 

Complex nature of accretion disks formation in DM halos must include both the radiation transfer \citep{luo18,ardaneh18}, the MHD physics \citep[e.g.,][]{begelman23,begelman23b,latif23a,latif23b}, and their combination \citep[e.g.,][]{jiang20}. Some of the important questions looming to be answered are as follows. Do accretion disks around the SMBHs in active galactic nuclei (AGN) differ from the pre-SMBH disks forming in direct collapse? How does the magnetic field is seeded in these disks? What is the main mechanism that amplifies the field within direct collapse accretion disks? What is the topology of the field close to the rotation axis of the disk, and is it suitable for launching a jet?

In this paper we analyze the results of high-resolution zoom-in cosmological simulations of direct collapse in the presence of very weak seed magnetic fields leading to the formation of a rotationally-supported disk during the pre-SMBH phase. We focus on the topology of the evolving magnetic field in the central $\sim 0.1$\,pc of the collapse, and analyse the instabilities associated with its growth. 

The outline of this work is as following. Section\,\ref{sec:numerics} deals with the numerical methods used, section\,\ref{sec:results} provides the results, which are discussed and summarized in sections\,\ref{sec:discuss} and \ref{sec:conclusions}.

\section{Numerical Methods and Simulation Setup}
\label{sec:numerics}

We conduct 3D cosmological simulations utilizing the Eulerian adaptive mesh refinement (AMR) code Enzo-2.6\footnote{https://github.com/enzo-project/enzo-dev/tree/gold-standard-v15} \citep{norman99,bryan14}. 
The gravitational interactions within collisionless systems are computed through a particle-mesh $N$-body approach \citep{hockney88}, which incorporates a multi-grid Poisson solver for self-gravitational computations.

The MHD equations are solved using the hyperbolic cleaning technique introduced by \citet{dedner02}. The primary integration is performed through the MUSCL 2nd order Runge-Kutta method \citep{vanleer79}. Additionally, the Poisson equation is solved for a divergence cleaning, by employing conjugate gradient with a 4-cell stencil. The local Harten-Lax-van Leer Riemann solver (HLL; \citet{harten83}) is utilized as a two-wave, three-state solver to compute fluxes at the cell boundaries. For specifying the interpolation scheme for the left and right states in the Riemann problem, a reconstruction is carried out using the piecewise linear method (PLM; \citet{vanleer79}).

In the MHD initial setup, the seed magnetic field strength is initialized at $\rm 10^{-15}\,G$ at the start of the simulation, $z = 199$. Previous studies have shown that regardless of the initial magnetic field intensity, the magnetic field converges towards an equipartition level when strong accretion shocks are present during the collapse phase \citep[e.g.,][]{latif23a}.

The gas chemistry is assumed to be  of the primordial composition, and we use the publicly available package GRACKLE-3.1.1\footnote{https://grackle.readthedocs.org/} \citep{bryan14,smith17} to follow the thermal and chemical evolution of the collapsing gas. GRACKLE is an open-source chemistry and radiative cooling/heating tool designed for use in numerical astrophysical simulations. 

The low resolution pathfinder simulation was conducted starting from redshift $z = 199$ and continued until $z = 10$, with cosmological initial conditions. It utilized a $1\,h^{-1}$\,Mpc comoving box with a root grid dimension of $128^3$, generated by the MUSIC package \citep{hahn11}, without the AMR and the baryonic component. The identification of a suitable DM halo was accomplished through the HOP group finder \citep{eisenstein98}. Subsequently a zoom-in DM halo has been generated with $1024^3$ effective resolution in DM and gas, centered on the selected halo position. The selected zoom-in region was sufficiently large to prevent interference from the high-mass, low-resolution DM particles. Within the zoom-in region, $\sim 10^7$ refined DM particles have been used, yielding an effective DM particle mass of about 99\,M$_{\odot}$. 

The baryon resolution was determined by the size of the grid cells, which are adaptively refined based on three criteria: the baryon mass, DM mass, and the Jeans length. If the gas or DM densities in a region of the simulation grid exceed $\rho_0 2^{(3+\alpha) l}$, where $\rho_0$ is the threshold density for refinement (assumed to be four times the mean density of the subgrid), the grid is refined by a factor of $2$ in length at the refinement level $l$, with $\alpha$ set to $-0.2$ for super-Lagrangian refinement \citep{bryan14}. To prevent failure in solving the MHD equations, the maximum CFL-implied time-step or Courant safety number is limited to $0.02$.

We impose the condition of at least 48 cells per Jeans length in our simulations, to avoid the artificial fragmentation \citep{truelove97}. Moreover, to sufficiently resolve the turbulent eddies and maintain the small-scale dynamo, the Jeans resolution of at least $32$ cells has been ensured during the entire course of simulations \citep{federrath11}. The maximum refinement level of $25$ has been used, which provides about $1.0\times10^{-5}$\,pc or 2\,AU physical resolution when this refinement level is reached. 

We introduce an exponential cut-off in the optically-thin cooling rate, $\Lambda$ (calculated by GRACKLE), above a predetermined threshold density $\rho_{\rm c}$. Opacity is introduced, represented by the optical depth $\tau$, which is determined by the local density and defined as $\tau = (\rho/\rho_{\rm c})^2$ \citep[e.g.,][]{hirano17}. When the local density exceeds $\rho_{\rm c}$, the total cooling rates are adjusted to $\Lambda \exp(-\tau)$. In our simulations, we choose $\rm \rho_{\rm c} = 10^{-10}\ g\,cm^{-3}$, a value derived from the density-temperature relationship identified in \citet{luo18} and \citet{ardaneh18}. As the density rises in the central regions, the introduction of an optical thickness leads to the formation of an optically thick region which mimics formation of a super-massive star (SMS) or a self-gravitating disk.

The Planck\,2015 data has been used to specify the cosmological parameters \citep{planck-collaboration16}: $\Omega_{\rm m}$ = 0.3089, $\Omega_\Lambda$ = 0.6911, $\Omega_{\rm b}$ = 0.04859, $\sigma_8$ = 0.8159, $n_{\rm s}$ = 0.9667, and the Hubble constant $h$ = 0.6774 in units of $100\,\rm km\,s^{-1}\,Mpc^{-1}$.

We use $R$ to abbreviate the spherical radii, and $r$ for cylindrical radii.  

\section{Results}
\label{sec:results}

\begin{figure}
\includegraphics[width=0.45\textwidth]{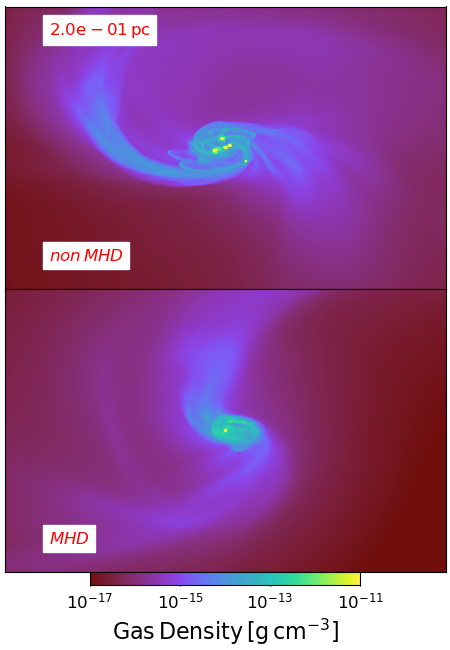}

    \caption{Face-on accretion disk projections of the gas density using snapshots at $t\sim 2.5$\,kyr. The width of each frame is $2\times 10^{-1}$\,pc. {\it Top:} accretion disk without the magnetic field. {\it Bottom:} same disk evolved with the seed magnetic field.}
    \label{fig:disk01}
\end{figure}  
 
The gravitational collapse happens at $z\sim 11$ in the run. At the beginning of the collapse, the DM halo has a virial mass of $\sim 5.2 \times 10^7\,M_\odot$, virial radius of $R_{\rm vir}\sim 1.0h^{-1}$kpc, and the cosmological spin parameter, $\uplambda\sim 0.035$. We have renormalized the time in our simulations to $t = 0$, when the gas collapse has reached the maximum refinement level. Time before $t = 0$ is considered being negative. The simulation has continued until $t\sim 2.5$\,kyr, to follow the central mass accumulation and analyze its morphology and other basic parameters, including the evolution of the magnetic field. 

\subsection{Formation of a pre-SMBH accretion disk}
\label{sec:disk}

The disk formation is preceded by two converging filaments that trigger a shock at $\sim 1$\,pc, which changes abruptly the direction of the angular momentum inside this region, and the emerging disk spin reflects this complex flow. Figure\,\ref{fig:disk01} displays the forming accretion disk and its structure on the larger, $\sim 0.1$\,pc scale. By this time, the disk exists inside $r\sim 10^{-2}$\,pc, and the spiral arms outside this radius extend for another decade to 0.1\,pc, gradually bending out of the disk plane. At the very center, due to a buildup of an optical depth and a reduced cooling, a central core starts to form.  By the end of the run the total mass within 0.1\,pc exceeds $10^4\,M_\odot$. More than 90\% of this mass belongs to the accretion disk outside the core. The DM fraction is less than 7\% inside this radius. Hence, the disk is self-gravitating.

This underlines an important difference between accretion disks around the SMBHs, which can be self-gravitating only at the large radii \citep[e.g.,][]{shlosman87,shlosman89a,shlosman89,goodman03}, and our pre-SMBH disks, which are self-gravitating at all radii, --- probably the only such objects existing in the history of the universe. Consequently, the physics of direct collapse accretion disks can differ profoundly from the former disks. Only some of these differences relevant to present work are discussed here. 

For comparison, we run the model with identical initial conditions but without introducing the seed magnetic field. The hydro solver and code implementation remain consistent across all comparison runs. Figure\,\ref{fig:disk01} shows the end products of both simulations. While the large scale structure of the non-MHD disk is very similar to the disk with the MHD, the small scale structure differs profoundly by triggering fragmentation. The fragments initially reside in the spiral arms, then spiral in towards the disk core and merge, but new fragments form continuously \citep[e.g.,][]{prole24}. Similar evolution of the fragments has been detected in \citet{shlosman16}. We conclude that the magnetic field, as expected, stabilizes the disk \citep[e.g.,][]{hopkins24b}, although \citet{latif23b} still obtain fragmentation in the magnetized disk.

As we show in section\,\ref{sec:disk}, the disk has been delineated in the outer part by the surface accretion shock and supported by magnetic field in the inner part. The disk thickness can be approximated well by $h\sim 5\times 10^{-3}$\,pc everywhere outside the core.  The decrease in the disk scaleheight in the core (where it is supported by the thermal pressure gradient) has resulted in response to a larger surface density.

We refrain from using the stability parameter $Q$ \citep{toomre64} because the disk is geometrically thick. As such it requires an exponential factor exp$(h/r)$ in definition of $Q$, resulting in $Q>>1$. The magnetic field acts as a stabilizing agent by further increasing the value of $Q$ by contributing the Alfvenic velocity, $v_{\rm A}$, in addition to the sound speed, $c_{\rm s}$, and turbulent velocity, $v_{\rm t}$. Indeed, no fragmentation has been detected anywhere in the MHD disk accordingly. 

The effective viscosity, $\alpha_{\rm eff}$, of a turbulent accretion disk has been defined in terms of the equatorial stress tensor, $w_{\rm r\phi} = \alpha_{\rm eff} P_{\rm th}$ \citep[e.g.,][]{shakura73}, where $P_{\rm th}$ is the thermal pressure, and can be extended by invoking contribution from the magnetic stress tensor \citep[e.g.,][]{shakura73,hawley95}. In the cylindrical coordinates, it can be written as $\alpha_{\rm eff} = < \rho v_\phi v_{\rm r}>/<\rho c_{\rm s}^2>$, where $v_\phi, v_{\rm r}$ and $c_{\rm s}$ are the gas tangential velocity, radial velocity and the sound speed, respectively, and we average, both over the scaleheight and azimuthally, to account for possible vertical correlations between the disk density profile and accretion velocity.

\begin{figure}
    \includegraphics[width=0.45\textwidth]{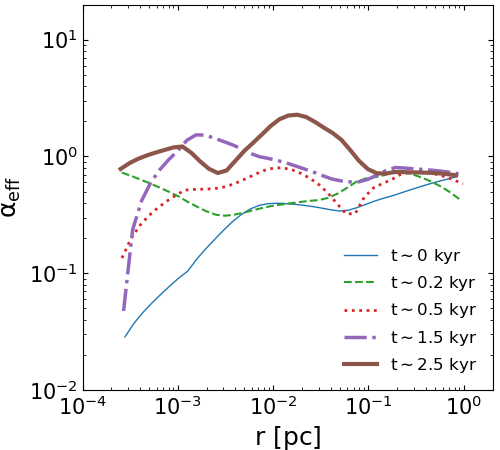}

    \caption{The effective viscosity, $\alpha_{\rm eff}$ (defined in the text), as a function of cylindrical radius $r$ in the accumulating accretion disk in the range of $r\sim 2\times 10^{-4}-1.0$\,pc, thus avoiding the core. $\alpha_{\rm eff}$ has been averaged over disk scaleheight, $h$, and azimuthally. }
    \label{fig:effviscosity}
\end{figure}

We display the effective viscosity parameter $\alpha_{\rm eff}$ in Figure\,\ref{fig:effviscosity}. As concluded by \citet{hawley95}, it increases with the field strength. For regular accretion disks around SMBHs, the main contribution to the angular momentum transfer are the magnetic and turbulent viscosity. But in self-gravitating MHD disks, i.e., when the central object has a negligible mass or is absent, the gravitational torques can contribute to the angular momentum transfer as well. This is exactly the case analyzed in this work. In the former disks, we expect $\alpha_{\rm eff} < 1$, but in the latter case, it can exceed unity.

Figure\,\ref{fig:effviscosity} shows $\alpha_{\rm eff}(r)$ in the range of $2\times 10^{-4}$\,pc --- 1\,pc. At early times, $\alpha_{\rm eff} < 1$, but at later times $\alpha_{\rm eff}\gtorder 1$ at some radii. We relate the latter abnormal values to the non-local gravitational viscosity\footnote{Note that magnetic viscosity can also be non-local \citep[e.g.,][]{blandford82}, although \citet{lesur21} has shown that non-local magnetic torques manifest themselves as a sink term rather than a viscous term.} \citep{shlosman90}. Note that the abnormally high values of effective viscosity are found outside $r\sim 10^{-2}$\,pc, where two open spiral arms are located, as shown in Figure\,\ref{fig:disk01}. Such spirals correspond to the Fourier mode $m=2$.

\begin{figure}
    \includegraphics[width=0.45\textwidth]{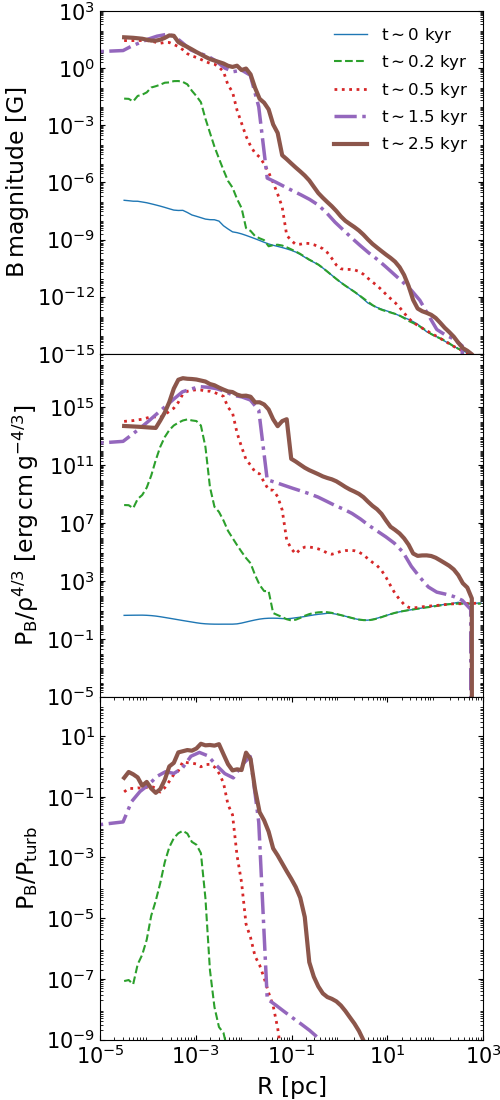}

    \caption{Amplification of magnetic field as function of the spherical radius $R$. Evolution of $B$ magnitude (top frame), $B^2/\rho^{-4/3}$ ratio (middle), and $P_{\rm B}/P_{\rm turb}$ (bottom) as a function of spherical radius $R$ at five representative times. Note that the $R$-dependence of the above quantities differs from their dependence in the accretion disk.}
    \label{fig:haloB}
\end{figure}    

 \begin{figure*}
    \includegraphics[width=0.98\textwidth]{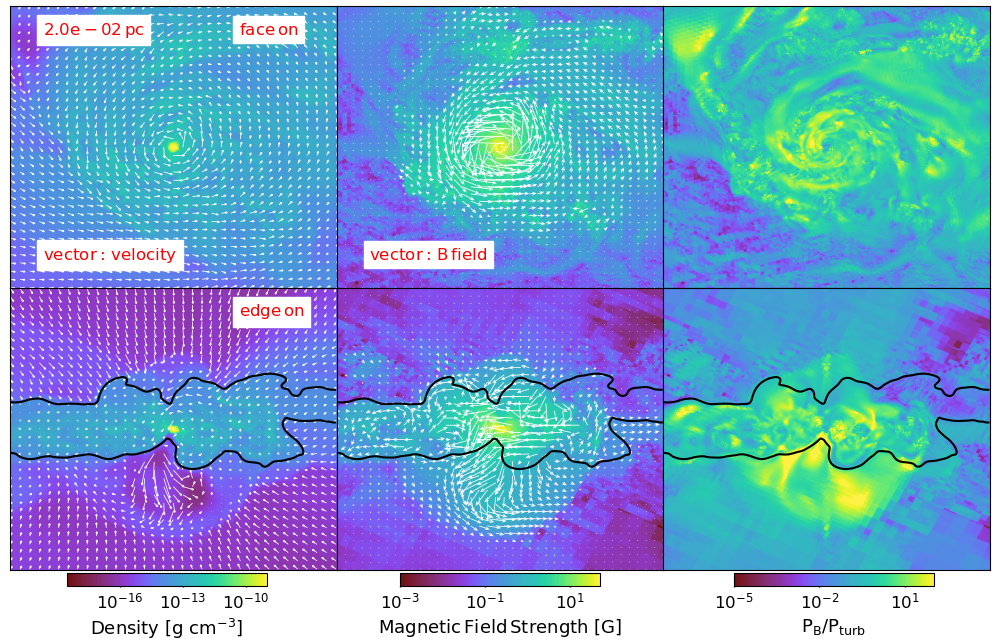}

    \caption{Face-on (top) and edge-on (bottom) disk slice view of the gas density (left row), magnetic field (central row), and the ratio of magnetic energy over turbulent energy, $\sim B^2/P_{\rm turb}$ (right row), at $t\sim 2.5$\,kyr. The velocity and magnetic field vectors are shown. The width of each frame is $2\times 10^{-2}$\,pc. The edge-on disk contours are abbreviated by solid black lines and result from a combination of the magnetic support and the surface shock by the accretion flow, as explained in the text.}
    \label{fig:diskSnapshot1}
\end{figure*}    

 \begin{figure*}
    \includegraphics[width=0.98\textwidth]{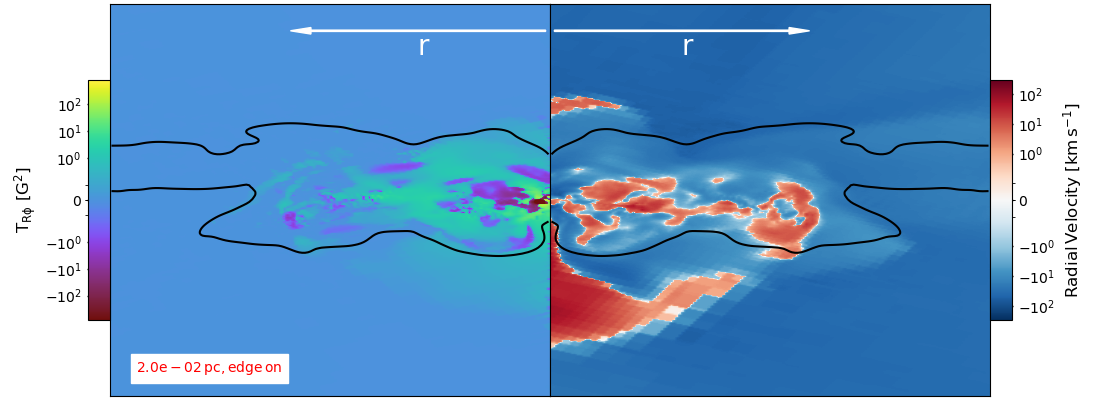}

    \caption{Edge-on right half-disk slice (direct and left-inverted), as in Figure\,\ref{fig:diskSnapshot1} at $t = 2.5$\,kyr. {\it Left frame:}  the radial Maxwell stress, $T_{\rm R\phi} = -B_{\rm R}B_\phi$. Close to the mid-plane, the negative stress shows the inflow of angular momentum and the mass outflow. {\it Right frame:} spherical radial velocity. Close to the disk mid-plane, the positive radial velocity shows an outflow and/or vortical motion.} 
    \label{fig:Mstress}
\end{figure*}

\begin{figure}
    \includegraphics[width=0.45\textwidth]{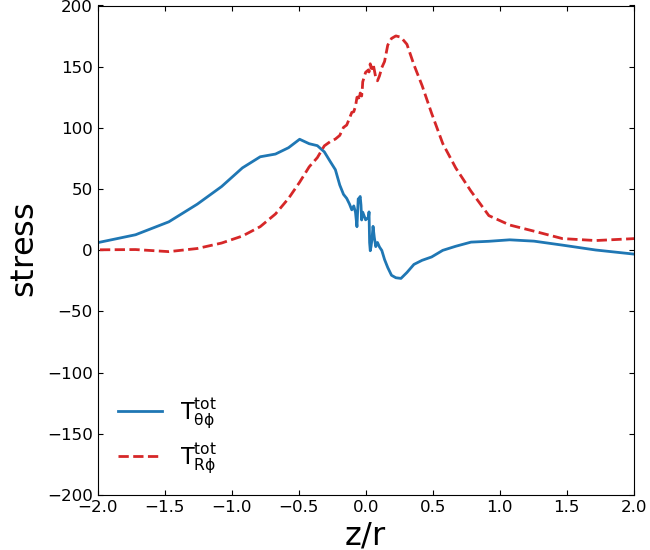}

    \caption{The total disk stresses, $T_{\rm \theta\phi}^{\rm tot}$ (solid blue line) and $T_{\rm R\phi}^{\rm tot}$ (dashed red line), which are the sum of the Maxwell and Reynolds stresses, as a function of $z/r$ at $t=2.5$\,kyr. The stresses have been calculated at representative $r=5\times 10^{-4}$\,pc and averaged azimuthally.}
    \label{fig:stress}
\end{figure}    

\subsection{Magnetic amplification in the pre-SMBH accretion disk}
\label{sec:magnetic}

Gravitational collapse within the central $\sim 0.1$\,pc attains a disky character. This transition is unrelated to the resolution, which is sufficient to resolve this characteristic radius by orders of magnitude, but it can depend on other parameters, such as the DM halo spin. 

From the kpc to pc scale, the magnetic field is amplified efficiently, well above what is expected from the compressional amplification, $B^2\sim \rho^{-4/3}$, as is evident in Figure\,\ref{fig:haloB}. The shocks and developed turbulence in the quasi-spherical collapse are responsible for this amplification. Over seven decades in $R$, the field has reached $\sim 100$\,G and leveled off at the center at $R\sim 10^{-4}$\,pc after $\sim 0.5$\,kyr. Over this range, $\rho^{4/3}$ has increased by about $\sim 18$ orders of magnitude. On the other hand, the $B^2/\rho^{4/3}$ ratio has increased to $\sim 10^{17}$ at $2\times 10^{-4}$\,pc, and declined a little at smaller radii at run end.  

The ratio of the magnetic field energy density to turbulent energy density, $P_{\rm turb}$, has increased by nine orders of magnitude from 10\,pc to $10^{-2}$\,pc, emphasizing the action of turbulence in the quasi-spherical collapse (see section\,\ref{sec:intro}).  Here, we define $P_{\rm turb} \sim \rho v_{\rm t}^2$, where $v_{\rm t}^2 = (v_{\rm r} - <v_{\rm r}>)^2 + (v_\phi - <v_\phi>)^2 + (v_{\rm z} - <v_{\rm z}>)^2$.

On the sub-pc scale, we demonstrate the accretion disk outline and its properties in Figure\,\ref{fig:diskSnapshot1}. The top row shows the face-on density, the magnetic field strength, and $P_{\rm B}/P_{\rm turb}$ ratio at the simulation end. Here $P_{\rm B}$ and $P_{\rm turb}$ denote the magnetic and turbulent pressure, respectively. Spiral arms are clearly visible in all three frames. The magnetic field is stronger in the spirals, and the magnetic energy density dominates there over the turbulent energy density. 

In the edge-on row in this Figure, the disk exhibits a same thickness with radius on the average. We have determined the thickness of the disk by outlining the surface shock produced by the accretion flow. As we show later in this section, the shock occurs when the accretion flow encounters the magnetically- or thermally-supported disk. The latter one even leads to a vertical compression of the disk.

What is not shown in Figure\,\ref{fig:diskSnapshot1}, is the time-dependent topology of the field. We observe that the $B$-field is heavily affected by the MHD turbulence. In particular, we observe vortices on various scales, even as large as $\sim 0.5h$. The vortical axes are both horizontal, i.e., pointing towards the observer in the edge-on disk, as well as parallel to the disk axis. This points to the disk field components experiencing large variations around the average values.  

In all three lower frames of Figure\,\ref{fig:diskSnapshot1}, the outflow along the disk rotation axis is visible. On the left frame, the outflow is pushing dense shells (more visible in the lower hemisphere). The outflow region can be also traced by the $B$-field structure. It is dominated strongly by the magnetic energy density over the turbulent energy density in the right frame. 

The important question is what drives the outflow along the polar axis of the accretion disk. While the dominance of the magnetic energy density over the turbulence hints to the magnetic force being the outflow driver, all dominant forces should be compared. Note that we ignore the radiation force in this simulation. We postpone this discussion to section\,\ref{sec:zresults}, where we analyze
the vertical structure of the disk.

The magnetic and turbulent states of the accretion disk, and the angular momentum transfer, can be characterized by invoking the Maxwell and Reynolds stresses. We have calculated the radial and polar (i.e., vertical) components of these stresses, $T_{\rm R\phi}$, $T_{\rm \theta\phi}$, $R_{\rm R\phi}$ and $R_{\rm \theta\phi}$, respectively, as well as the total stresses,  and display $T_{\rm R\phi}$ in Figure\,\ref{fig:Mstress}. The positive $T_{\rm R\phi}$, which determines the time derivative of the specific angular momentum per unit volume, $j_{\rm z}\sim \rho v_\phi R$, leads to the mass {\it inflow} and angular momentum flux outwards.

To analyze the the magnetic field and accretion velocity in the right frame of Figure\,\ref{fig:diskSnapshot1}, we plot the radial Maxwell stress, $T_{\rm R\phi} = -B_{\rm R} B_\phi$, responsible for the radial torque in the disk and the associated radial velocity, in Figure\,\ref{fig:Mstress}, both for the same half-disk. The blue shades (in the right frame) represent the inflow radial velocities and the green shades (in the left frame) to positive radial Maxwell stress leading to the inflow. The positive velocities and the negative stress correspond to the outflow, which is found in the equatorial region of the disk and in the polar outflows. While the polar outflow is related to the expanding bubbles which presumably will be squashed by the accretion flow (as we argue in section\,\ref{sec:discuss}), the equatorial positive velocities are similar to those discussed in the literature \citep[e.g.,][]{igumenshchev96,suzuki14,zhu18,mishra20}. 

Close to the disk mid-plane, the perturbations in the disk make the inner region of the disc unstable by producing vortices, especially if the polarity of the field is inverted \citep[e.g.,][]{abramowicz92,igumenshchev96, kluzniak20}. These vortices redistribute angular momentum, and can contribute to the equatorial outflows. But the dominant contribution to the outflow comes from the negative gradient of the total vertical stress, i.e., $T_{\rm \theta\phi}^{\rm tot} = T_{\rm \theta\phi} + R_{\rm \theta\phi}$, as shown in Figure\,\ref{fig:stress}.  This implies that the angular momentum flux is directed towards the disk midplane, and the mass flux there is directed outwards. The turbulent contribution in our disk can be substantial due to the time-dependent inflow.

In Figure\,\ref{fig:stress} we show the vertical profiles of the azimuthally-averaged total radial and vertical (polar) stresses, $T_{\rm R\phi}^{\rm tot}$ and $T_{\rm \theta\phi}^{\rm tot}$, at the representative $r\sim 5\times 10^{-4}$\,pc at the end of the simulation. The total stresses are the sum of the Maxwell and Reynolds stresses defined above. The radial stress is about a factor of two higher than the vertical stress. The vertical profile of the radial stress is nearly always positive, therefore ensuring the net inflow in the disk region. While the vertical stress changes its sign close to the disk midplane --- this behavior is governed by the components of the magnetic field, as discussed above.

\begin{figure}
     \includegraphics[width=0.45\textwidth]{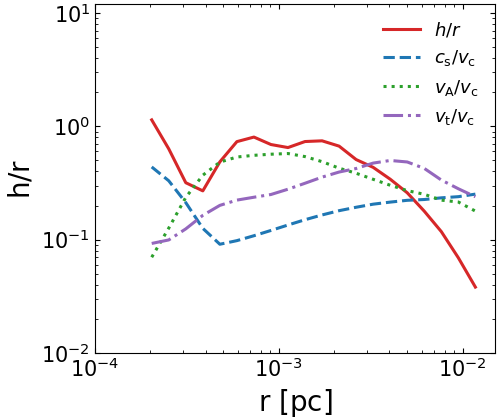}

    \caption{The disk thickness over cylindrical radius, $h/r$, at $t=2.5$\,kyr. The red solid line is adopted from Figure\,\ref{fig:diskSnapshot1}; the blue dashed line corresponds to $c_{\rm s}/v_{\rm c}$, the dotted line is $v_{\rm A}/v_{\rm c}$, and the dash-dotted line is $v_{\rm t}/v_{\rm c}$, where $v_{\rm c}$ is the Keplerian velocity. The disk is supported vertically by the magnetic field between $r\sim 3\times 10^{-4}-5\times 10^{-3}$\,pc, by the thermal pressure inside $r\sim 3\times 10^{-4}$\,pc, and is compressed by the accretion flow outside $r\sim 5\times 10^{-3}$\, pc. All the parameters have been averaged azimuthally and over the disk thickness. }
    \label{fig:thickness}
\end{figure}

Now, we can provide a stricter definition of the vertical thickness of accretion disk, $h$, normalized by $r$. Figure\,\ref{fig:thickness} displays the disk thickness delineated in Figure\,\ref{fig:diskSnapshot1} as a red solid line. Over a wide range, $r\sim 3\times 10^{-4}-5\times 10^{-3}$\,pc, it follows closely the line describing the magnetic support. For larger radii, the disk appears to be compressed by the accretion flow. Consequently, when averaging over $h$, we use $h\sim 5\times 10^{-3}$\,pc which closely follows the red line in this Figure. 

\begin{figure}
   \includegraphics[width=0.45\textwidth]{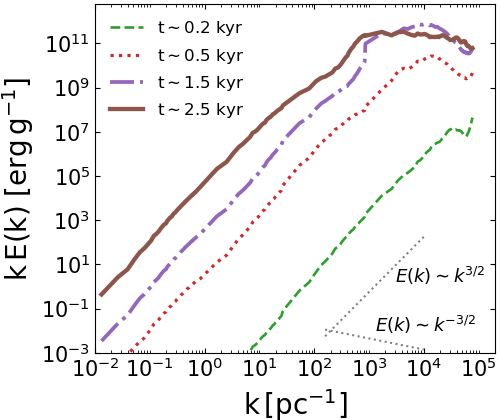}

    \caption{Power spectra of the MHD turbulence at representative times. The Kolmogorov spectrum of energy cascade with the spectrum $E(k)\sim k^{-5/3}$ with the wave-number $k$ has been reproduced (not shown) for the inertial range of $10^{-2}\,{\rm pc^{-1}}\ltorder k \ltorder 10^5\,{\rm pc^{-1}}$. The spectral energy density profile evolution confirms that $E(k) \sim k^{3/2}$ is injected at largest scales (here $\sim 100$\,pc). The turnover happens at later times and progresses to $ k\sim 10^3\,{\rm pc^{-1}}$. Formation of $k^{-3/2}$ branch at later times is also visible. 
    }
    \label{fig:PowerSp}
\end{figure}

\begin{figure}
    \includegraphics[width=0.45\textwidth]{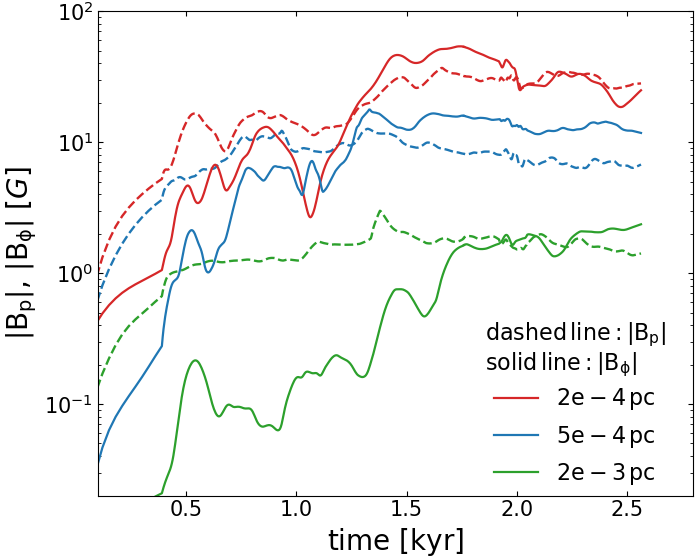}

    \caption{Growth of magnetic radial poloidal $B_{\rm p}$ and toroidal $B_\phi$ fields as function of time. Profiles show the fields at cylindrical radius $\sim 2\times 10^{-4}$\,pc, $\sim 5\times 10^{-4}$\,pc, and $\sim 2\times 10^{-3}$\,pc --- all values are averaged over the disk thickness, $h$, and azimuthally.}
    \label{fig:Bevol}
\end{figure}

\begin{figure*}
    \includegraphics[width=0.99\textwidth]{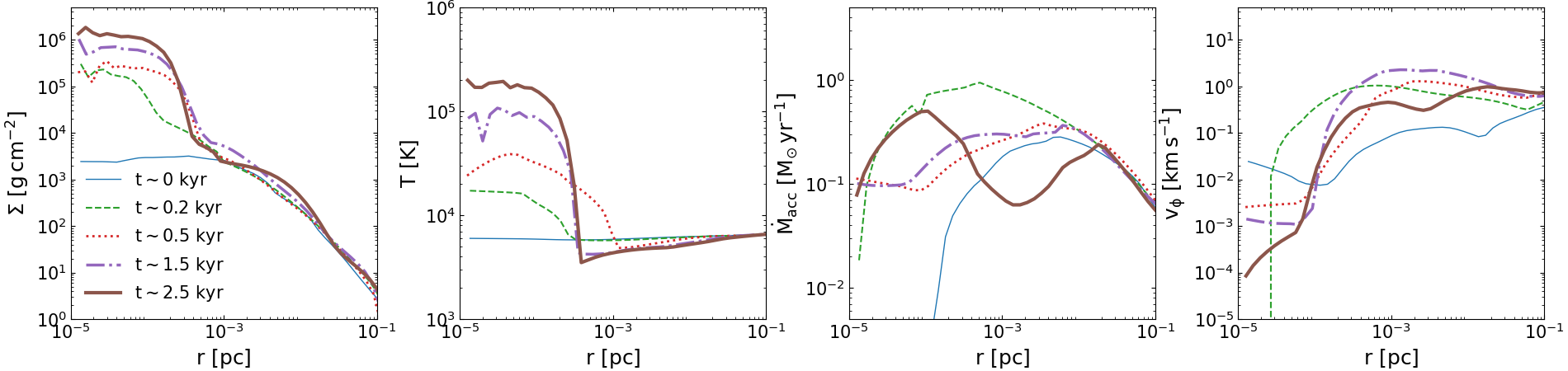}

  \caption{Evolution of the profiles of surface density $\Sigma$, temperature $T$, accretion rate $\dot M_{\rm acc}$, and the tangential velocity $v_{\phi}$, as function of cylindrical radius $r$ in the accumulating accretion disk within $r\sim 0.1$\,pc, at different times. All the variables have been averaged over the disk thickness, $h$, and azimuthally. }
    \label{fig:diskFaceOn}
\end{figure*}

\begin{figure*}
    \includegraphics[width=0.99\textwidth]{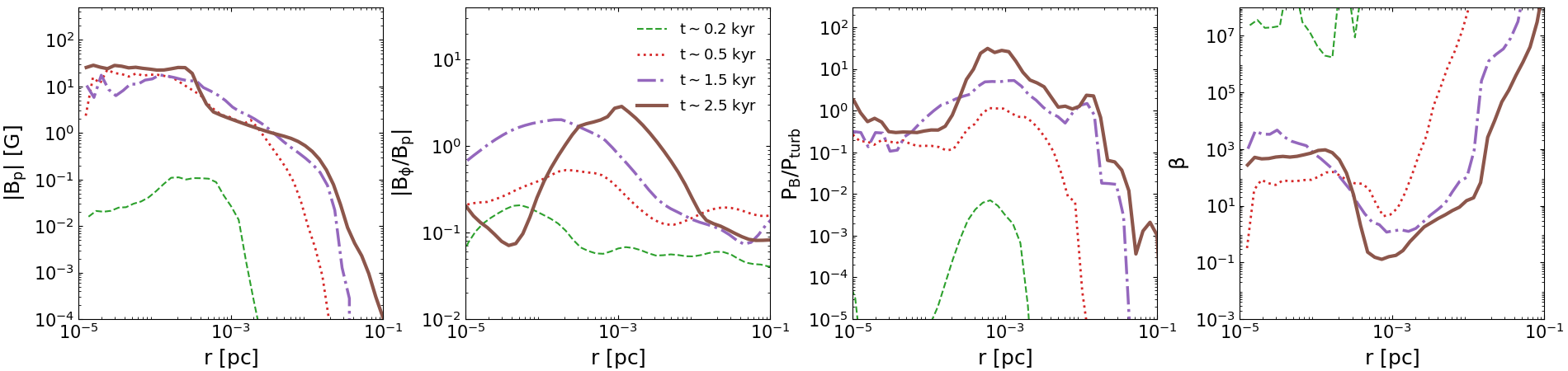}

  \caption{Evolution of the profiles of (radial) poloidal field $B_{\rm p}$, ratio $B_{\rm \phi}/B_{\rm p}$, ratio $P_{\rm B}/P_{\rm turb}$, and $\beta$ as function of cylindrical radius $r$ in the accumulating accretion disk within $r\sim 0.1$\,pc, at different times. All the variables have been averaged over the disk thickness, $h$, and azimuthally.}
    \label{fig:diskFaceOn2}
\end{figure*}

The magnetic field has been amplified both during the quasi-spherical collapse and subsequently in the accretion disk (see Figs.\,\ref{fig:haloB} and \ref{fig:Bevol}-\ref{fig:diskFaceOn2}). This small scale dynamo, as formulated in \citet{kazantsev68} and \citet{kraichnan68}, has been tested in Figure\,\ref{fig:PowerSp} along the lines given in \citet{kulsrud92}. The turbulent cascade has been investigated for the spatial range of $\sim 10^{-5}-10^2$\,pc, for a few representatives times. The field amplification process has been found active initially for the whole range, with the turbulent energy spectrum $E(k)\sim k^{-5/3}$ reproducing the Kolmogorov spectrum. On larger scales, where the collapse is quasi-spherical, it is the small-scale dynamo that acts to amplify the field, while on smaller scales, where the flow becomes disky, the field is amplified by the dynamo associated with the MRI, which excites turbulence. In the latter case, the torques are dominated by the magnetic Maxwell stress over Reynolds stress, and the effective viscosity increases with the field strength \citep[e.g.,][]{hawley95}, as indeed is observed here (see Figure\,\ref{fig:Mstress} and below). 

The energy spectrum $E(k)$ in Figure\,\ref{fig:PowerSp} exhibits the \citet{kazantsev68} energy spectrum of $\sim k^{3/2}$ and dominates throughout the $k$-range at earlier times. Starting with $t=0.5$\,kyr and later, the spectrum saturates at large $k\sim 10^4\,{\rm pc^{-1}}$ due to a strong $B$ field, and the saturation gradually moves towards smaller $k$. At the highest $k$, the spectrum turns over and displays the \citet{iroshnikov63} slope of $k^{-3/2}$. 

The Kazantsev spectrum $E(k)\sim k^{3/2}$ means that the small-scale turbulent dynamo operates and cascades the energy to large scales and increases the power $E(k)$ there. The appearance of the Iroshnikov spectrum $E(k)\sim k^{-3/2}$ at later times reflects the shutting off the small dynamo on these scales. We relate this termination of the small-scale dynamo to the Parker instability which is triggered at this time, as discussed in section\,\ref{sec:discuss}. At this time, the peak of $E(k)$ lies around $10^3-10^4\,{\rm pc^{-1}}$, which separates the two slopes.


Here, we emphasize an important point about the field amplification in direct collapse simulations with the seed field. During the formation of the accretion disk within the central $\sim 0.01$\,pc, it is seeded by a relatively strong magnetic field which has been amplified during the quasi-spherical stage of the collapse. This field however is below equipartition, with $B_{\rm p}\sim 10^{-4}$\,G, $B_\phi/B_{\rm p}\sim 0.1$, $P_{\rm B}/P_{\rm turb}\sim 10^{-6}$ and $\beta = P_{\rm th}/P_{\rm B}\sim 10^7$. These appear as initial values to be amplified by the MRI in the disk. Hence, we do not seed the initial field in the disk, as is done in local and global simulations of isolated disks, where vertical poloidal field is artificially seeded \citep[e.g.,][]{mishra20}.

To ensure that the MRI sustains the turbulence, we analyze the parameter $Q_{\rm MRI}$ to assess the resolution quality of the MRI. Here, $Q_{\rm MRI} \equiv \lambda_{\rm MRI}/\Delta x$, with $\lambda_{\rm MRI} = B/\sqrt{4\pi\rho} \Omega$, $\Delta x$ being the local cell size, and $\Omega$ representing the rotational frequency \citep[eg.,][]{noble2010,hawley2011}. We have found $Q_{\rm MRI}$ within the disk to be approximately 

\begin{equation}
Q_{\rm MRI}\sim 300\bigg(\frac{B}{1\,G}\bigg) \bigg(\frac{\rho}{10^{-13}\,{\rm g\,cm^{-3}}}\bigg)^{-0.5}\bigg(\frac{\Omega}{10^{-10}\,s^{-1}}\bigg)^{-1}\bigg(\frac{\Delta x}{10^{-5}\,{\rm pc}}\bigg)^{-1} ,
\end{equation}
which exceeds the empirical minimum value $Q_{\rm MRI, min}\sim 20$, required to characterize MRI-driven turbulence \citep{hawley2011}. 

Next, we focus on the inner disk region, inside $\sim 10^{-2}$\,pc, and follow the evolution of the magnetic field in the disk mid-plane --- both poloidal and toroidal fields there. Figure\,\ref{fig:Bevol} displays the evolution of the field components averaged over the disk thickness, until $t\sim 2.5$\,kyr. All components grow exponentially at early times, $t\ltorder 0.5$\,kyr, then continue the linear growth. The poloidal field, $B_{\rm p}$, grows and saturates from outside in. That means, the outer component grows first and saturates at higher value. The inner components grow slower and saturate at lower values. The toroidal components of the field, $B_\phi$, grow slower but saturate close to the poloidal components. The saturation means that the field has reached the near-equipartition value.

Unlike accretion disks around SMBHs, the direct collapse disks modeled here are low-temperature, magnetized disks with $c_{\rm s}$, $v_{\rm t}$, $v_{\rm A}$ and $v_{\rm c}$ velocities which differ relatively little, $\sim 10$, from each other {\it within the disk}. Here $v_{\rm c}$ is the Keplerian velocity. This factor is important in order to understand the underlying physics of these disks. The MRI saturation in our models agrees with the \citet{pessah05} formalism, when $B^2\sim 4\pi \rho c_{\rm s} v_{\rm c}$. Under the above equipartition condition, the field saturation value exceeds $B^2\sim 4\pi \rho c_{\rm s}^2$ by a relatively small factor of few$\times 10$ only.  

We analyze the magnetic evolution in cylindrical coordinates and express the important parameters in the same geometry --- namely, using cylindrical radius $r$, surface density $\Sigma(r)$, temperature $T(r)$, accretion rate $\dot M_{\rm acc}(r)$, and the tangential velocity $v_\phi(r)$ of the disky flow --- all averaged over the disk thickness (see Figure\,\ref{fig:diskFaceOn}). The model has been run for $\sim 2.5$\,kyr after it reached the maximal refinement level. By this time, the core has reached high volume density at the center, $\sim 10^{-9}\,{\rm g\,cm^{-3}}$, and hence surpassed the critical density, above which the suppressed cooling rate has been imposed (section\,\ref{sec:numerics}). This cooling rate is lower than that for the optically-thin gas.  

The disky region can be roughly considered to lie in the range of $r\sim 2\times 10^{-4}-10^{-1}$\,pc, and the central core lies at smaller radii. While outer slope is $\Sigma\sim r^{-3/2}$ and does not change with time, the core surface density is steadily rising with time, remaining flat. Disk temperature stays flat as well at the floor temperature. On the other hand, the core temperature rises above $\sim 10^5$\,K --- a possible signature of evolution towards the hydrostatic equilibrium.

The accretion rate in the disk fluctuates, but remains flat overall, between $\dot M_{\rm acc}\sim 0.1-1\,M_\odot\,{\rm yr^{-1}}$. Because of the combination of a spherical and disk accretion inside $r\sim 0.1$\,pc, the disk $\dot M$ seems to increase towards smaller $r$ as the quasi-spherical mass flux is added to the disk via surface shock.   On the other hand, inside the core the accretion rate drops. The tangential velocity, $v_\phi$ is flat in the disk, fluctuating between $\sim 0.1-{\rm few}\,{\rm km\,s^{-1}}$. Inside the core, it drops, reflecting the trend towards hydrostatic equilibrium there --- but note that radiation force which is omitted here can change this equilibrium dramatically.

Figure\,\ref{fig:diskFaceOn2} displays the radial profiles of various magnetic properties averaged over the accretion disk thickness and azimuthally. The poloidal field $B_{\rm p}$ grows steadily with smaller $r$ and flattens inside the core. At the same time, the ratio $B_{\rm z}/B_{\rm p}$ stays flat between $\sim 0.1-1$, while $B_\phi/B_{\rm p}\gtorder 1$ and peaks in the disk at later times. Similar behavior we observe for $P_{\rm B}/P_{\rm turb}$. This behavior finds its explanation in the $\beta$ profile.  The accretion disk is heavily magnetized, with $\beta<1$, while for the core, $\beta>1$.

Figure\,\ref{fig:diskFaceOn} shows that the surface density saturates inside $r\sim 2\times 10^{-4}$\,pc, forming a spinning core. Essentially being the result of the increasing optical depth and the inability of the gas to cool, it can be studied as a possible progenitor of a growing supermassive star (SMS) \citep{begelman10,woods21,herrington23,woods24}. The basic parameters of this core should be understood as the result of a quasi-adiabatic equation of state approximation and the reduced cooling rate above $\rho_{\rm c}$ used in this simulation. 

\begin{figure}
    \includegraphics[width=0.45\textwidth]{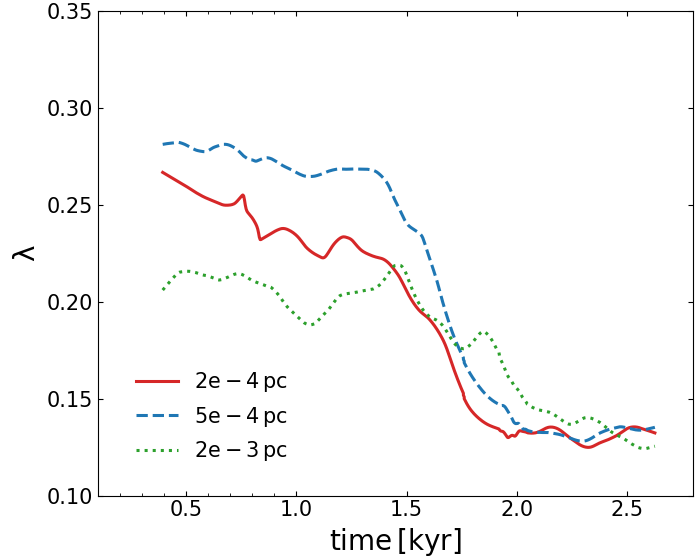}
    \caption{
    The spin parameter $\uplambda = J/J_{\rm max}$ within three representative radii, including the core radius of $\sim 2\times 10^{-4}$\,pc, as function of time. 
     }
    \label{fig:spin}
\end{figure}

The core spins down with time by a factor of $\sim 2$ during the simulation, but remains rotationally supported, as shown in Figure\,\ref{fig:spin}. We define its spin as $\uplambda = J/J_{\rm max}$, where $J$ is the total angular momentum within certain radius $r$, and $J_{\rm max}$ is its maximal (Keplerian) angular momentum.  At $t=0$, within the core radius of $2\times 10^{-4}$\,pc, $\uplambda\sim 0.27$, and at the end of the run it is $\sim 0.13$. The spin down is accelerating after $t\sim 1.5$\,kyr, when, as we show later, the magnetic field becomes dynamically important and hence magnetic torques are amplified.

\subsection{Vertical gradients of gas and magnetic field parameters}
\label{sec:zresults}

\begin{figure*}
    \includegraphics[width=0.99\textwidth]{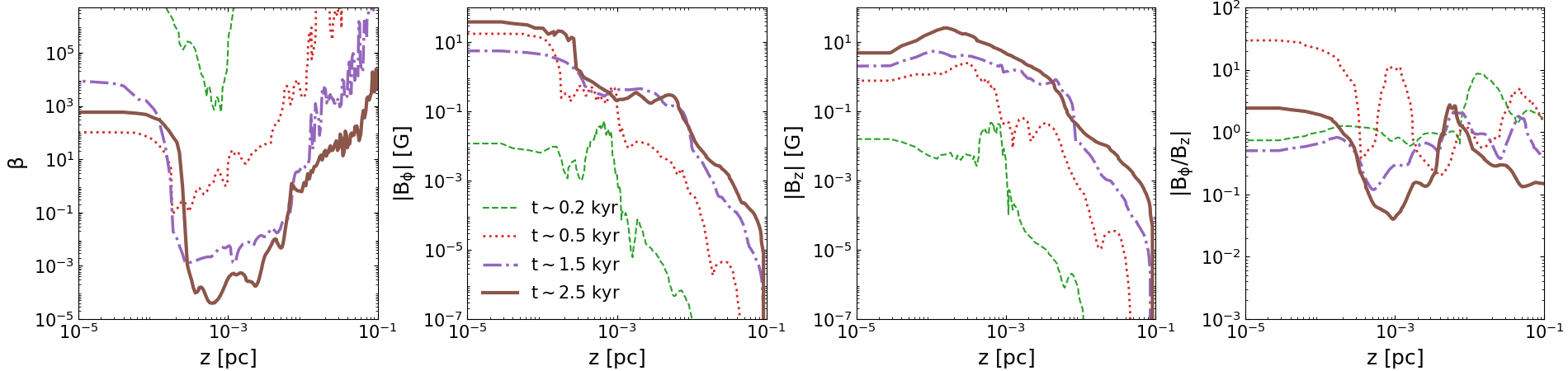}

    \caption{Evolution of magnetic field profiles along the polar axis $-z$ (i.e., the lower hemisphere shown in Figure\,\ref{fig:diskSnapshot1})  of the accretion disk: $\beta$, toroidal field $B_{\phi}$, vertical poloidal $B_{\rm z}$ field, and the ratio $B_{\phi}/B_{\rm z}$ at four characteristic times. The measurements have been taken at $r = 0$.}
    \label{fig:diskZprofile}
\end{figure*}    

One of the important aspects of direct collapse is the buildup of magnetic fields in the central region, especially within the accretion disk. We, therefore, turn to analyzing the dependence of various parameters in the disk along its polar axis. To have a more general view of the vertical distribution of density and magnetic properties, we sample these profiles at $r=0$ and at a number of azimuthal points at $r=10^{-3}$\,pc. 

As we have discussed in the previous section, the core extends to $r\sim 10^{-4}$\,pc as can be observed from Figure\,\ref{fig:diskFaceOn} (left frame). The vertical extension of the core is visible also along the polar axis, and the plateau has developed within $z\sim 2\times 10^{-4}$\,pc, which reflects the density profiles inside the core --- $\rho(z)$ remains flat and apparently saturates after $t\sim 1.5$\,kyr. The reason for this saturation is related to the outflow observed in Figure\,\ref{fig:diskSnapshot1}.

The disk scaleheight evolves with time in tandem with the strength of the magnetic field in the disk. As we show in section\,\ref{sec:discuss}, instability in the magnetic field drives some of the disk material along its rotation axis, spreading it along the disk surface, thus increasing the disk thickness. As a result, the disk vertical structure is related to the evolution of the magnetic field structure \citep[e.g.,][]{mishra20}.

The vertical structure of the MHD accretion disks in the presence of a central compact object has been investigated numerically in the local and global approximations neglecting the gas self-gravity and assuming the vertical hydrostatic equilibrium \citep[e.g.,][]{mishra20}. This allowed to dissect the accretion flow and observe the equatorial counter-flow and the higher altitude incoming flow --- these are detected here as well (e.g., Figs.\,\ref{fig:diskSnapshot1} and \ref{fig:Mstress}). Moreover, an addition of the radiation flux would lead to the vertical convection in the disk \citep[][]{jiang20}.

Evolution of $\beta$ in Figure\,\ref{fig:diskZprofile} reveals the buildup of the magnetic energy density {\it above} the core, with $\beta << 1$ along the polar axis after $t\sim 0.5$\,kyr. The very low values of $\beta\sim {\rm few}\times 10^{-5}$ within the outflow cavity are remarcable because it is much lower than $\beta\sim 0.1-1$ in the disk mid-plane. As the disk has fully formed inside $r\sim {\rm few}\times 10^{-2}$\,pc by this time, this points to the field expulsion from the disk --- we address this phenomenon below and in section\,\ref{sec:discuss}. 

While the toroidal field strength is flat within the core and falls outside it, the vertical poloidal field $B_{\rm z}$ is slowly increasing and decreasing, overall staying flat up to $z\sim 2\times 10^{-2}$\,pc. Then it falls sharply as well. But the ratio $B_\phi/B_{\rm z}$ decreases to $\sim {\rm few}\times 10^{-2}$ above the disk inside the expanding bubble, and stays above 0.1 up to $z\sim 0.1$\,pc. Remembering that the disk thickness is $h\sim 5\times 10^{-3}$\,pc, we find that $B_{\rm z}$ is generally higher above the disk than in its interior at the end of the run. This is also true for the core.

To obtain a more complete picture of magnetic field evolution above the disk, we have sampled the field values along the polar axis but at $r\sim 10^{-3}$\,pc, i.e. away from the core (not shown here).  We find that the disk is far from being axisymmetric, because the mass density, $\beta$, and the ratio $B_\phi/B_{\rm z}$ all vary by a factor of $\sim 10$ at the end of the simulation. But the minimal value of $\beta$ is nearly constant. The ratio $B_\phi/B_{\rm z}$ varies by a factor of $\sim 5$.   

The evolution of magnetic field above the disk is limited by the simulation time. But the trend is clear --- buildup of vertical poloidal and toroidal fields along the polar axis, above the the disk.

\begin{figure*}
    
    \includegraphics[width=0.99\textwidth]{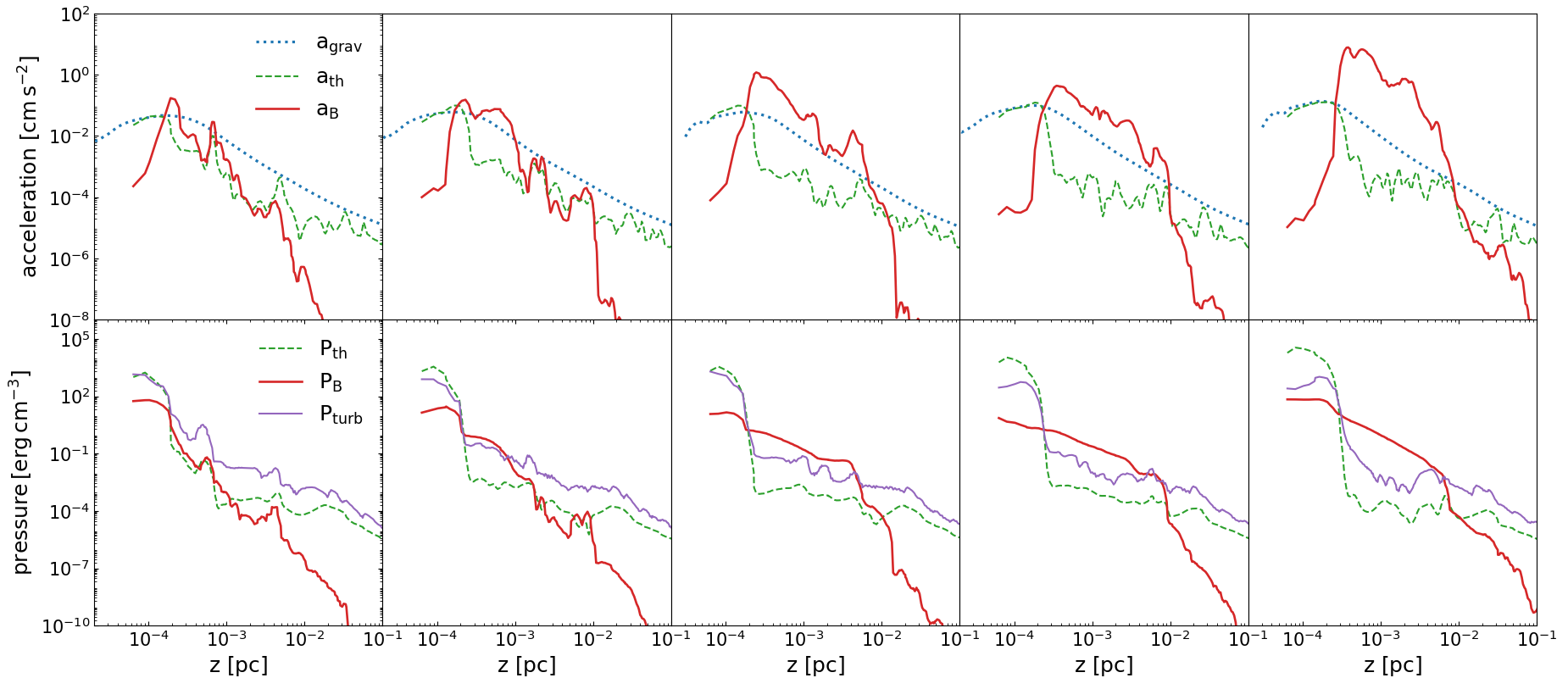}

    \caption{{\it Top panel:} Dominant acceleration profiles, along the polar axis of accretion disk starting with $t=0.5$\,kyr and for $t=0.5$\,kyr increments --- gravitational acceleration, $a_{\rm grav}$, acceleration based on the thermal pressure gradient, $a_{\rm th}$, and acceleration based on the gradient of magnetic pressure, $a_{\rm B}$. 
    {\it Bottom panel:} thermal ($P_{\rm th}$), magnetic ($P_{\rm B}$) and turbulent ($P_{\rm turb}$) pressure profiles along the polar axis of the accretion disk at the same times. All the variables have been measured at $r = 0$. }
    \label{fig:accele}
\end{figure*}    

\begin{figure*}
 \begin{center}
\includegraphics[width=0.9\textwidth]{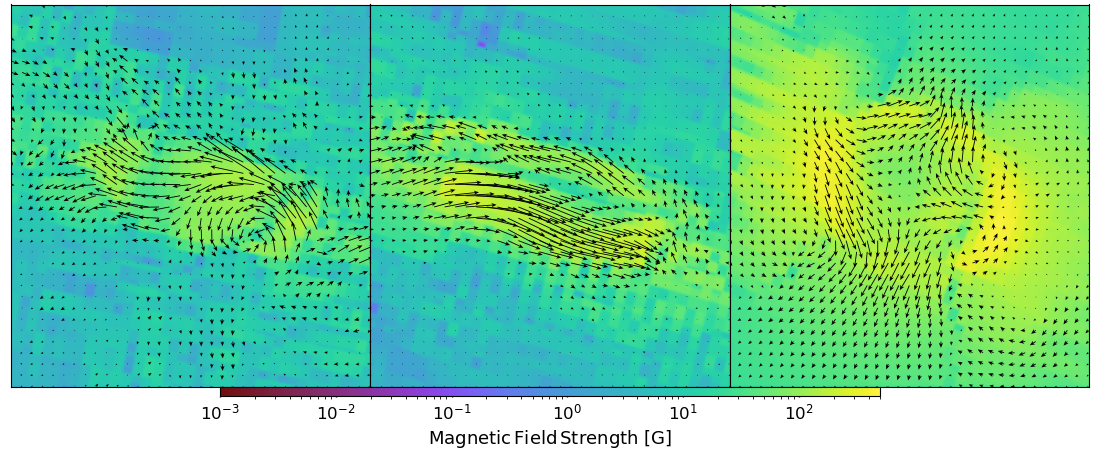}
 \end{center}
    \caption{Edge-on disk slices within the central $10^{-3}$\,pc. The color palette shows the magnetic field strength, and the arrows display the direction of the total magnetic field ---here dominated by the toroidal field. The left frame taken at earlier times shows
    a turbulent disk with two dominant vortices (upper left and lower right) which are oriented at random angles with respect to the disk rotation. The middle frame display a growing toroidal field with inverted polarity at the mid-plane, which experiences a minimum around the equatorial plane. The right frame taken at the end of the run, $t=2.5$\,kyr, shows a buckling toroidal field being expelled from the disk. The orientation of the slices is exactly the same as in Figure\,\ref{fig:diskSnapshot1}.   }
    \label{fig:diskSnapshot2}
\end{figure*}    

To understand the dominant forces acting above the accretion disk, and in particular in the outflow region, we calculate the associated accelerations based on the gravity and thermal and magnetic pressure gradients along the disk polar axis, namely, $a_{\rm grav}$, $a_{\rm th}$ and $a_{\rm B}$ respectively (Figure\,\ref{fig:accele}, top frames). The radiation field is ignored in our simulations and the perturbations can be considered as adiabatic. The gravitational acceleration depends on the surface density of the disk, $\sim \pi G\Sigma$, and the overall mass distribution in the central region. Outside the core, $\rho\sim R^{-2}$ (Figure\,\ref{fig:haloB}), which agrees with the observed $a_{\rm grav}\sim R^{-1}$. Inside the core it declines.  We calculate the two additional accelerations and thermal, magnetic and turbulent pressure distributions above the disk at $t=0.5$\,kyr with the time increments of 0.5\,kyr (shown in Figure\,\ref{fig:accele}, top and bottom frames, respectively). 

Initially, the gravitational acceleration slightly dominates at all $z$, while the magnetic and thermal accelerations are comparable but smaller. The magnetic acceleration and magnetic pressure start to dominate after $t\sim 1.5$\,kyr, and this trend extends to larger $z$ with time. By the end of the run, the magnetic acceleration dominates between $z\sim 2\times 10^{-4}$\,pc and $z\sim 10^{-2}$\,pc, and this region will increase further based on the outflow velocity reaching $\sim 80\,{\rm km\,s^{-1}}$ within the expanding cavity which drives the slower shock, as an expanding hemisphere.
Note that the three pressure distributions differ little from each other inside the core.

This outflow region can be clearly observed in Figures\,\ref{fig:diskSnapshot1} and \ref{fig:diskSnapshot2}, inside $z\sim 10^{-2}$\,pc, where it is dominated by the magnetic pressure over the turbulent pressure. On the opposing side of the disk, the same expansion occurs, although shifted in time. The backflow towards the disk can be observed as well (in the first lower frame of Figure\,\ref{fig:diskSnapshot1}), as discussed in \citet{kato98}. 

As a next step, we analyze the vertical disk structure in order to determine the evolving topology of the magnetic field and how it is related to the generated outflow. Figure\,\ref{fig:diskSnapshot2} shows the edge-on slice of the accretion disk within the central $10^{-3}$\,pc oriented the same way as in Figure\,\ref{fig:diskSnapshot1}.  The arrows display the orientation of the magnetic field, and the color palette corresponds to the $B$-field strength.

The left snapshot, taken around $t\sim 0.5-1$\,kyr, displays the forming turbulent disk with the MHD vortices. Two of these vortices can be observed, at the upper left and lower right of the disk. As we mentioned before, these vortices' axes do not seem to correlate with the disk angular momentum vector. In principle, this vortical motion can contribute to variability in the outflow velocities in the disk midplane, which is dominated by a negative total stress shown in Figures\,\ref{fig:Mstress} and especially in Figure\,\ref{fig:stress}. This variability which is superposed on a steady phenomena \citep[e.g.,][]{mishra20}, can be observed here in Figure\,\ref{fig:diskSnapshot2}. Quantifying this variability is a subject of a future work.

The middle frame in Figure\,\ref{fig:diskSnapshot2} has been taken around $t\sim 1.5-2$\,kyr, and the third frame at the end of the run. Note that the field lines display an opposite polarity above and below the disk mid-plane, i.e., reversal with respect to the mid-plane. This anti-symmetry is particularly favorable to the field instability as we discuss in the next section. Moreover, in the middle frame, the field is oriented parallel to the mid-plane in the inner disk, it shows the beginning of expulsion of the field from the disk --- the field is pushed away from the mid-plane and begins to buckle. With the decrease of the field in the mid-plane region, it could be dominated by the thermal and or turbulent pressure.

In the right frame of Figure\,\ref{fig:diskSnapshot2}, away from the polar axis, the magnetic field is dominated by the poloidal field which points inwards. Closer to the polar axis, the field is dominated by the {\it vertical} poloidal component. And the transition region between both regions displays the gradual bending or buckling of the field direction outwards along the polar axis.  

Our comparison with \citet{mishra20} work should be taken with the grain of salt, as the model conditions differ fundamentally. While Mishra et al. deals with steady state flow and not accounting for the gas self-gravity, our model is far from being in steady state, both radially and vertically, is not strictly axisymmetric, and accounts fully for the changing gravity. Probably, the most significant difference comes from the innermost boundary condition, which includes the SMBH in the Mishra et al. work.

\section{Discussion}
\label{sec:discuss}

We have modeled direct collapse towards the SMBH seed within DM halo using high resolution zoom-in  cosmological simulation in the presence of a seed magnetic field of $\sim 10^{-15}$\,G. The magnetic field has been amplified initially during a quasi-spherical stage of direct collapse by the gas compression, but mostly by the generated turbulence. A self-gravitating accretion disk has formed inside $\sim 0.1$\,pc and has been growing. After the simulation reached the maximal resolution of $10^{-5}$\,pc or 2\,AU and its density went above the critical one changing its equation of state to quasi-adiabatic one, it has been continued for additional 2.5\,kyr.  

The field has been growing inside the disk, from the field deposited there by the quasi-spherical accretion flow, reaching a near-equipartition value. The accretion disk became strongly magnetized with time, with $\beta\sim 10^2$ averaged in the core, dropping  to  $\beta\sim 0.1$ at larger radii in the disk. This evolution has been fast, and $\beta$ dropped below unity already after $\sim 1.5$\,kyr. No fragmentation has been detected in the MHD disk, while a similar non-MHD disk model has shown a multiple and ongoing fragmentation.

As a by-product of this evolution in the accretion disk, we have detected an outflow along the polar axis. The outflow is weakly collimated and drives a shock wave away and along the disk surface, as a growing bubble. The outflow region is dominated by the magnetic field, and the ratio of magnetic-to-turbulent energy density ratio being much larger than unity. We have followed the evolution of the field above the disk and found that $\beta$ there drops even faster and more profoundly than in the disk mid-plane, reaching the value of nearly ${\rm few}\times 10^{-5}$, which is $\sim 3$ orders of magnitude below the disk mid-plane value. This effect points strongly in the direction of magnetic field being expelled from the accretion disk. 

Furthermore, the vertical poloidal field strength, $B_{\rm z}$, has remained flat up $z\sim 2\times 10^{-2}$\,pc. The ratio $B_\phi/B_{\rm z}\sim 0.1-1$ along the polar axis, meaning that the toroidal field there has reached the value of $\sim 1$\,G. Finally, we find that the gradient of magnetic energy density along the $z$-axis is the driver of the detected outflow.

What is the fate of this outflow in the presence of the ongoing accretion? We did not follow the evolution of the expanding bubbles long enough due to the termination of this run and the relatively mild outflow velocities, but compared the kinetic pressure of the outflow with that of the accretion flow which joins the disk via surface shock. In both cases, the kinetic pressure is $\sim \rho v^2$, where $\rho$ is the postshock density delineating the bubble, or in the accretion flow near the disk, and $v$ is the shock outflow velocity or the accretion velocities. We estimate the kinetic pressure of the former as $\sim 0.1\,{\rm erg\,cm^{-3}}$, while that of the latter as  $\sim 10^5\,{\rm erg\,cm^{-3}}$. Hence, the inevitable conclusion is that the outflow will be stopped by the accretion flow. Although one should remember that we have avoided the radiation pressure which can make the difference \citep[e.g.,][]{luo23}.

So the bubbles will be squashed and flattened against the disk surface by the ongoing accretion flow, still contributing to the disk geometrical thickening. In this most probable scenario, the magnetic pressure within this layer will only increase due to compression and remain much higher than the near-equipartition field close to the disk mid-plane, thus enhancing the existing configuration of a magnetized disk with a magnetosphere. We avoid calling the magnetosphere as a corona, as no temperature rise is distinguishable there. Within the magnetosphere, we find that $v_{\rm A} > c_{\rm s}$. This situation can change of course when a radiation transfer is introduced.

\subsection{The Parker instability}
\label{sec:parker}

To explain the observed evolution of the field lines, we refer to the buoyancy force which develops as the field grows in the disk. The related instability is the Parker instability \citep{parker66}. This instability is characterized with the field line deformation, and the magnetic tension acts to stabilize the instability. 

As we observe in the second frame of Figure\,\ref{fig:diskSnapshot2}, we detect the near disappearance of the $B$-field in the midplane region, which is expected due to the field polarity reversal there. Under these conditions, the diffusion of the opposite fields towards the (midplane) boundary will naturally lead to the field reduction there, a well known effect.

The field lines are characteristically bending upwards, buckle and expand. As the disk angular velocity has a radial gradient, the rising field lines will also twist around the vertical axis. This evolution  resembles separation of the magnetic fields into individual flux tubes --- a process which requires additional force to counteract the isotropic magnetic pressure. The additional force can only be the gravity. The third frame in this Figure exhibits a clear buckling of the field lines upwards, away from the disk mid-plane, and is associated with a strong growth of the field along the polar axis --- a nonlinear stage of the Parker instability. 

Conditions for the Parker instability in accretion disks has been analyzed both analytically and numerically for the case of a central object responsible for the gravity \citep[e.g.,][]{horiuchi88,matsumoto88}. The current disk is self-gravitating and lacks the central object. 

The instability develops when its associated wavelength $L$ along $r$, over which the $B$ field line is raised, fulfills the condition,  

\begin{equation}
L \gtorder \bigg(\frac{1}{\delta} + \frac{v_{\rm A}^2}{c_{\rm s}^2} ,  \bigg)^{1/2} h
\end{equation}
where $\delta\equiv (P_{\rm B}/P_{\rm th})$, and $v_{\rm A}$ is the Alfvenic speed. $h$ is the isothermal scale-height in the disk. For shorter wavelengths, the magnetic tension, $\sim B^2/l$, stabilizes the instability (where $l$ is radius of curvature of the field). While for much longer wavelength the growth rate of instability is very high. The anti-symmetric field orientation with respect to the disk mid-plane, i.e., as in Figure\,\ref{fig:diskSnapshot2}, acts favorably to increase in the growth rate of the Parker instability \citep{horiuchi88}. We note that the field polarity we detect in the disk can change abruptly, apparently due to substantial perturbations of the field by horizontal vortices discussed earlier. However, investigating this effect is beyond the scope of the current work.

Figure\,\ref{fig:diskSnapshot2} shows that at the onset of instability $\delta\sim 1$, $v_{\rm A}\sim c_{\rm s}$ and $h\sim 1.5\times 10^{-4}$\,pc. Hence, the condition on $L$ is $L\gtorder 2\times 10^{-4}$\,pc, which agrees well with Figure\,\ref{fig:diskSnapshot2}, and requires that the unstable wavelength should exceed the core size, as observed.

\subsection{Field amplification during vertical acceleration}
\label{sec:coriolis}

Additional by-product of the nonlinear Parker instability is the amplification of the toroidal and vertical poloidal fields above the disk, which we observe in Figure\,\ref{fig:diskZprofile}, where the ratio $B_\phi/B_{\rm z}$ lies within the range of $\sim 0.1-{\rm few}$, as a function of $z$. 

In the present work, which neglects radiation forces, we have concluded that the outflow will be squashed against the surface of the disk by the accretion flow. We conjecture about the fate of this outflow and the field if radiation force would provide additional push, as we have observed in \citet{luo23}, where the magnetic fields have been omitted. A strong radiation force along the $z$-axis acting on the fully ionized gas would provide a drag on the toroidal magnetic lines which have already experienced the Parker instability, and enhance the buoyancy. 

Because the underlying disk is subject to Keplerian shear, this amplifies the upward dent in the field lines. The upward motion during Parker instability is similar to that of a convection cell. Such an upwelling is known as cyclonic --- the gas spins as it rises, and will spin faster than its surroundings because of the converging flow in the dent and conservation of the angular momentum, leading to the production of a meridional field, $B_{\rm z}$ \citep[e.g.,][]{parker55}. Overall, this process resembles convection, but we conjecture that in the presence of a radiation force the symmetry in upwelling-downwelling is broken. 

Returning to the description of the magnetic flux tubes used in the previous section, the production of the poloidal $B_{\rm z}$ field happens in order to compensate for the shear inside the tube. The 3-D AMR numerical simulations of disks with the Keplerian shear have verified that under these conditions a Coriolis force, $\sim 2\Omega \rho v_{\rm r}$, is exerted on the buoyant flux tube, acting to separate the tube sides, while twisting of the field stabilizes the tube against deformation due to the magnetic surface tension \citep[e.g.,][]{ziegler01}. Here $\rho$ and $v_{\rm r}$ are the density inside the expanding tube and its radial expansion velocity. In our simulations, the same argument can be applied by replacing the flux tube description by the expanding bubble which spins around the disk axis producing a coherent helicity.

Hence, the role of the Coriolis force in the above process of uplifting the toroidal field by Parker instability is to produce a coherent helicity, which is the source of an $\alpha$-effect in the $\alpha$-$\Omega$ dynamo in a differentially rotating disk. The helical motion twists the toroidal field and amplifies the $B_{\rm z}$ component \citep[e.g.,][]{pariev01}.  

\subsection{Comparison with the literature}
\label{sec:comparison}

We have compared our results with \citet{hopkins24b}, although the latter work deals with accretion disks around quasar-type SMBHs. So, majority of the differences originate because of differences in physical systems. Two fundamental differences between these models are (1) the innermost boundary, which is determined by the SMBH in the Hopkins et al. model, and (2) the near-equipartition saturation of the $B$-field, which agrees well with the \citet{pessah05} formalism of a strongly magnetized MRI.   

The similarities between our models include the following.

\begin{itemize}
\item We also do not observe fragmentation in the accretion flow in the presence of the $B$-field.
\item We also obtain a magnetically-dominated disk.
\item Our $P_{\rm B}/P_{\rm turb}$ is also ranging $0.1-10$.
\item We agree that the toroidal component becomes the dominant one.
\item We agree that the disk is magnetically supported in the large fraction of the disk, with $h/r\sim 0.1-1$.
\item We agree and show that the accretion is driven by a combination of Maxwell and Reynolds stresses.
\end{itemize}

At the same time, we find that

\begin{itemize}
\item Our $\beta$ is substantially higher.
\item We find a break in the $B$-field profile at the disk outer edge.
\item We conclude, based on evidence, that the field has been amplified by the strongly magnetized version of the MRI, and not by the flux-freezing. The MRI has saturated when reaching the predicted critical value of \citet{pessah05}.
\item We observe the Parker instability in the field which appears to limit the growth of field in the disk.
\end{itemize}

The source of these differences is partly based on the physics of accretion disks around an SMBH and the direct collapse disks, and will require further work.

\section{Conclusions}
\label{sec:conclusions}

In summary, direct collapse in DM halos at high redshifts with a primordial composition gas in the presence of a magnetic field seed leads to the formation of a self-gravitating geometrically-thick accretion disk within $\sim 0.1$\,pc. Whether this evolution results in the formation of a central supermassive star or continues the disky collapse until the formation of the horizon is beyond the scope of this work. We use high-resolution zoom-in cosmological simulations with the maximal resolution of $10^{-5}$\,pc, or 2\,AU. The seed magnetic field of $10^{-15}$\,G has been introduced within the computational box at $z=199$, and has been amplified during the quasi-spherical collapse within DM halo by orders of magnitude, but remained well under equipartition when the disk forms. Hence no additional seed field is required, compared to modeling isolated disks. This field has been amplified by the highly magnetized version of the MRI \citep{pessah05} to near-equipartition, when $B^2\sim 4\pi \rho c_{\rm s} v_{\rm c}$.  The magnetized disk does not show fragmentation, while a comparison non-MHD disk exhibits continuously forming fragments. As the MRI has switched off around equipartition field strength, the field lines experience the buoyancy instability which leads to the bending and buckling of the field lines vertically, i.e., the field experiences the nonlinear Parker instability. We show that this instability triggers an outflow driven by the magnetic field gradient. 

The outflow produces two expanding and strongly magnetized bubbles with $\beta<<1$, whose  evolution was only briefly followed here. Based on the bubble and accretion flow kinematics, we estimate that, in the absence of radiation force, expansion of these bubbles will be terminated by the kinetic pressure of the accretion flow, and they will be squashed against the disk surface. This evolution leads to the formation of a magnetized disk with $\beta\sim 10^{-1} - 10^2$, which is sandwiched by its magnetosphere with $\beta << 1$.

Overall, a pre-SMBH accretion disk forming due to direct collapse within a DM halo differs from an AGN disk around the SMBH in a number of ways. First, it is fully self-gravitating at all radii. Second, its structure is time-dependent and non-axisymmetric due to the inflow via the outer boundary, which is variable and highly turbulent. Third, while the disk magnetic field is amplified successfully by the MRI, the amplification process is also very time-dependent and disk is perturbed by vortical motions on scales even comparable to the disk thickness, having random orientation of their spin. Fourth, the near-equipartition field is subject to the Parker instability which leads to the formation of the disk magnetosphere and may lead to the formation of collimated jets under additional forces, e.g., the radiation force which is ignored in this model. Fifth, we have detected equatorial regions in the disk subject to a negative gradient of the total stress, which drives an equatorial outflow, sandwiched by an inward accretion flow. Finally, the absence of an SMBH in the disk modifies conditions in the central region, resulting in different thermodynamic and MHD conditions there. 

\section*{Acknowledgements}
All the analysis has been conducted using yt \citep{turk11}, http://yt-project.org/. I.S. is grateful to Mitch Begelman for discussions on various topics relevant to this work. We thank Marek Abramowicz for his comments about Figure\,5. Y.L. acknowledges the support from the NSFC grants No. 12273031. I.S. acknowledges the hospitality of KITP at UCSB where part of this research has been conducted under the NSF under Grant No. NSF PHY-1748958. I.S. is grateful for generous support from the International Joint Research Promotion Program at Osaka University. The simulations were carried out at the National Supercomputer Center in Tianjin, performed on TianHe-1A. 

\bibliography{ms}{} 

\begin{thebibliography}{}
\expandafter\ifx\csname natexlab\endcsname\relax\def\natexlab#1{#1}\fi
\providecommand{\url}[1]{\href{#1}{#1}}
\providecommand{\dodoi}[1]{doi:~\href{http://doi.org/#1}{\nolinkurl{#1}}}
\providecommand{\doeprint}[1]{\href{http://ascl.net/#1}{\nolinkurl{http://ascl.net/#1}}}
\providecommand{\doarXiv}[1]{\href{https://arxiv.org/abs/#1}{\nolinkurl{https://arxiv.org/abs/#1}}}

\bibitem[{{Abramowicz} {et~al.}(1992){Abramowicz}, {Lanza}, {Spiegel}, \&
  {Szuszkiewicz}}]{abramowicz92}
{Abramowicz}, M.~A., {Lanza}, A., {Spiegel}, E.~A., \& {Szuszkiewicz}, E. 1992,
  \nat, 356, 41, \dodoi{10.1038/356041a0}

\bibitem[{{Ardaneh} {et~al.}(2018){Ardaneh}, {Luo}, {Shlosman}, {Nagamine},
  {Wise}, \& {Begelman}}]{ardaneh18}
{Ardaneh}, K., {Luo}, Y., {Shlosman}, I., {et~al.} 2018, \mnras, 479, 2277,
  \dodoi{10.1093/mnras/sty1657}

\bibitem[{{Ba{\~n}ados} {et~al.}(2018){Ba{\~n}ados}, {Venemans},
  {Mazzucchelli}, {Farina}, {Walter}, {Wang}, {Decarli}, {Stern}, {Fan},
  {Davies}, {Hennawi}, {Simcoe}, {Turner}, {Rix}, {Yang}, {Kelson}, {Rudie}, \&
  {Winters}}]{banados18}
{Ba{\~n}ados}, E., {Venemans}, B.~P., {Mazzucchelli}, C., {et~al.} 2018, \nat,
  553, 473, \dodoi{10.1038/nature25180}

\bibitem[{{Balbus} \& {Hawley}(1991)}]{balbus91}
{Balbus}, S.~A., \& {Hawley}, J.~F. 1991, \apj, 376, 214,
  \dodoi{10.1086/170270}

\bibitem[{{Balbus} \& {Hawley}(1998)}]{balbus98}
---. 1998, Reviews of Modern Physics, 70, 1, \dodoi{10.1103/RevModPhys.70.1}

\bibitem[{{Begelman}(2010)}]{begelman10}
{Begelman}, M.~C. 2010, \mnras, 402, 673,
  \dodoi{10.1111/j.1365-2966.2009.15916.x}

\bibitem[{{Begelman} \& {Armitage}(2023)}]{begelman23b}
{Begelman}, M.~C., \& {Armitage}, P.~J. 2023, \mnras, 521, 5952,
  \dodoi{10.1093/mnras/stad914}

\bibitem[{{Begelman} \& {Shlosman}(2009)}]{begelman09}
{Begelman}, M.~C., \& {Shlosman}, I. 2009, \apj, 702, L5,
  \dodoi{10.1088/0004-637X/702/1/L5}

\bibitem[{{Begelman} \& {Silk}(2023)}]{begelman23}
{Begelman}, M.~C., \& {Silk}, J. 2023, \mnras, 526, L94,
  \dodoi{10.1093/mnrasl/slad124}

\bibitem[{{Bhowmick} {et~al.}(2024){Bhowmick}, {Blecha}, {Torrey},
  {Weinberger}, {Kelley}, {Vogelsberger}, {Hernquist}, \&
  {Somerville}}]{bhowmick24}
{Bhowmick}, A.~K., {Blecha}, L., {Torrey}, P., {et~al.} 2024, \mnras, 529,
  3768, \dodoi{10.1093/mnras/stae780}

\bibitem[{{Blandford} \& {Payne}(1982)}]{blandford82}
{Blandford}, R.~D., \& {Payne}, D.~G. 1982, \mnras, 199, 883,
  \dodoi{10.1093/mnras/199.4.883}

\bibitem[{{Bogd{\'a}n} {et~al.}(2024){Bogd{\'a}n}, {Goulding}, {Natarajan},
  {Kov{\'a}cs}, {Tremblay}, {Chadayammuri}, {Volonteri}, {Kraft}, {Forman},
  {Jones}, {Churazov}, \& {Zhuravleva}}]{bogdan23}
{Bogd{\'a}n}, {\'A}., {Goulding}, A.~D., {Natarajan}, P., {et~al.} 2024, NatAs,
  8, 126, \dodoi{10.1038/s41550-023-02111-9}

\bibitem[{{Bosman} {et~al.}(2024){Bosman}, {{\'A}lvarez-M{\'a}rquez}, {Colina},
  {Walter}, {Alonso-Herrero}, {Ward}, {{\~A}-stlin}, {Greve}, {Wright}, {Bik},
  {Boogaard}, {Caputi}, {Costantin}, {Eckart}, {Garc{\'\i}a-Mar{\'\i}n},
  {Gillman}, {Hjorth}, {Iani}, {Ilbert}, {Jermann}, {Labiano}, {Langeroodi},
  {Pei{\ss}ker}, {Rinaldi}, {Topinka}, {van der Werf}, {G{\"u}del}, {Henning},
  {Lagage}, {Ray}, {van Dishoeck}, \& {Vandenbussche}}]{bosman23}
{Bosman}, S. E.~I., {{\'A}lvarez-M{\'a}rquez}, J., {Colina}, L., {et~al.} 2024,
  NatAs, 8, 1054, \dodoi{10.1038/s41550-024-02273-0}

\bibitem[{{Brandenburg} {et~al.}(1996){Brandenburg}, {Jennings}, {Nordlund},
  {Rieutord}, {Stein}, \& {Tuominen}}]{brandenburg96}
{Brandenburg}, A., {Jennings}, R.~L., {Nordlund}, {\r{A}}., {et~al.} 1996,
  Journal of Fluid Mechanics, 306, 325, \dodoi{10.1017/S0022112096001322}

\bibitem[{{Brandenburg} \& {Subramanian}(2005)}]{brandenburg05}
{Brandenburg}, A., \& {Subramanian}, K. 2005, \physrep, 417, 1,
  \dodoi{10.1016/j.physrep.2005.06.005}

\bibitem[{{Bryan} {et~al.}(2014){Bryan}, {Norman}, {O'Shea}, {Abel}, {Wise},
  {Turk}, {Reynolds}, {Collins}, {Wang}, {Skillman}, {Smith}, {Harkness},
  {Bordner}, {Kim}, {Kuhlen}, {Xu}, {Goldbaum}, {Hummels}, {Kritsuk}, {Tasker},
  {Skory}, {Simpson}, {Hahn}, {Oishi}, {So}, {Zhao}, {Cen}, {Li}, \& {Enzo
  Collaboration}}]{bryan14}
{Bryan}, G.~L., {Norman}, M.~L., {O'Shea}, B.~W., {et~al.} 2014, The
  Astrophysical Journal Supplement Series, 211, 19,
  \dodoi{10.1088/0067-0049/211/2/19}

\bibitem[{{Chandrasekhar}(1960)}]{chandrasekhar60}
{Chandrasekhar}, S. 1960, Proceedings of the National Academy of Science, 46,
  253, \dodoi{10.1073/pnas.46.2.253}

\bibitem[{{Choi} {et~al.}(2013){Choi}, {Shlosman}, \& {Begelman}}]{choi13}
{Choi}, J.-H., {Shlosman}, I., \& {Begelman}, M.~C. 2013, \apj, 774, 149,
  \dodoi{10.1088/0004-637X/774/2/149}

\bibitem[{{Choi} {et~al.}(2015){Choi}, {Shlosman}, \& {Begelman}}]{choi15}
---. 2015, \mnras, 450, 4411, \dodoi{10.1093/mnras/stv694}

\bibitem[{{Dedner} {et~al.}(2002){Dedner}, {Kemm}, {Kr{\"o}ner}, {Munz},
  {Schnitzer}, \& {Wesenberg}}]{dedner02}
{Dedner}, A., {Kemm}, F., {Kr{\"o}ner}, D., {et~al.} 2002, Journal of
  Computational Physics, 175, 645, \dodoi{10.1006/jcph.2001.6961}

\bibitem[{{Eisenstein} \& {Hut}(1998)}]{eisenstein98}
{Eisenstein}, D.~J., \& {Hut}, P. 1998, \apj, 498, 137, \dodoi{10.1086/305535}

\bibitem[{{Emmering} {et~al.}(1992){Emmering}, {Blandford}, \&
  {Shlosman}}]{emmering92}
{Emmering}, R.~T., {Blandford}, R.~D., \& {Shlosman}, I. 1992, \apj, 385, 460,
  \dodoi{10.1086/170955}

\bibitem[{{Federrath} {et~al.}(2011){Federrath}, {Sur}, {Schleicher},
  {Banerjee}, \& {Klessen}}]{federrath11}
{Federrath}, C., {Sur}, S., {Schleicher}, D. R.~G., {Banerjee}, R., \&
  {Klessen}, R.~S. 2011, \apj, 731, 62, \dodoi{10.1088/0004-637X/731/1/62}

\bibitem[{{Goodman}(2003)}]{goodman03}
{Goodman}, J. 2003, \mnras, 339, 937, \dodoi{10.1046/j.1365-8711.2003.06241.x}

\bibitem[{{Hahn} \& {Abel}(2011)}]{hahn11}
{Hahn}, O., \& {Abel}, T. 2011, \mnras, 415, 2101,
  \dodoi{10.1111/j.1365-2966.2011.18820.x}

\bibitem[{Harten {et~al.}(1983)Harten, Lax, \& Leer}]{harten83}
Harten, A., Lax, P.~D., \& Leer, B.~v. 1983, SIAM Review, 25, 35,
  \dodoi{10.1137/1025002}

\bibitem[{{Hawley} {et~al.}(1995){Hawley}, {Gammie}, \& {Balbus}}]{hawley95}
{Hawley}, J.~F., {Gammie}, C.~F., \& {Balbus}, S.~A. 1995, \apj, 440, 742,
  \dodoi{10.1086/175311}

\bibitem[{{Hawley} {et~al.}(2011){Hawley}, {Guan}, \& {Krolik}}]{hawley2011}
{Hawley}, J.~F., {Guan}, X., \& {Krolik}, J.~H. 2011, ApJ, 738, 84,
  \dodoi{10.1088/0004-637X/738/1/84}

\bibitem[{{Herrington} {et~al.}(2023){Herrington}, {Whalen}, \&
  {Woods}}]{herrington23}
{Herrington}, N.~P., {Whalen}, D.~J., \& {Woods}, T.~E. 2023, MNRAS, 521, 463,
  \dodoi{10.1093/mnras/stad572}

\bibitem[{{Hirano} {et~al.}(2017){Hirano}, {Hosokawa}, {Yoshida}, \&
  {Kuiper}}]{hirano17}
{Hirano}, S., {Hosokawa}, T., {Yoshida}, N., \& {Kuiper}, R. 2017, Science,
  357, 1375, \dodoi{10.1126/science.aai9119}

\bibitem[{{Hockney} \& {Eastwood}(1988)}]{hockney88}
{Hockney}, R.~W., \& {Eastwood}, J.~W. 1988, {Computer simulation using
  particles}

\bibitem[{{Hopkins} {et~al.}(2024){Hopkins}, {Squire}, {Su}, {Steinwandel},
  {Kremer}, {Shi}, {Grudic}, {Wellons}, {Faucher-Giguere}, {Angles-Alcazar},
  {Murray}, \& {Quataert}}]{hopkins24b}
{Hopkins}, P.~F., {Squire}, J., {Su}, K.-Y., {et~al.} 2024, The Open Journal of
  Astrophysics, 7, 19, \dodoi{10.21105/astro.2310.04506}

\bibitem[{{Horiuchi} {et~al.}(1988){Horiuchi}, {Matsumoto}, {Hanawa}, \&
  {Shibata}}]{horiuchi88}
{Horiuchi}, T., {Matsumoto}, R., {Hanawa}, T., \& {Shibata}, K. 1988, \pasj,
  40, 147

\bibitem[{{Igumenshchev} {et~al.}(1996){Igumenshchev}, {Chen}, \&
  {Abramowicz}}]{igumenshchev96}
{Igumenshchev}, I.~V., {Chen}, X., \& {Abramowicz}, M.~A. 1996, \mnras, 278,
  236, \dodoi{10.1093/mnras/278.1.236}

\bibitem[{{Inayoshi} {et~al.}(2020){Inayoshi}, {Visbal}, \&
  {Haiman}}]{inayoshi20}
{Inayoshi}, K., {Visbal}, E., \& {Haiman}, Z. 2020, \araa, 58, 27,
  \dodoi{10.1146/annurev-astro-120419-014455}

\bibitem[{{Iroshnikov}(1963)}]{iroshnikov63}
{Iroshnikov}, P.~S. 1963, \azh, 40, 742

\bibitem[{{Jiang} \& {Blaes}(2020)}]{jiang20}
{Jiang}, Y.-F., \& {Blaes}, O. 2020, \apj, 900, 25,
  \dodoi{10.3847/1538-4357/aba4b7}

\bibitem[{{Kato} {et~al.}(2008){Kato}, {Fukue}, \& {Mineshige}}]{kato98}
{Kato}, S., {Fukue}, J., \& {Mineshige}, S. 2008, {Black-Hole Accretion Disks
  --- Towards a New Paradigm ---}

\bibitem[{{Kazantsev}(1968)}]{kazantsev68}
{Kazantsev}, A.~P. 1968, Soviet Journal of Experimental and Theoretical
  Physics, 26, 1031

\bibitem[{{Kluzniak} \& {Kita}(2000)}]{kluzniak20}
{Kluzniak}, W., \& {Kita}, D. 2000, arXiv e-prints, astro,
  \dodoi{10.48550/arXiv.astro-ph/0006266}

\bibitem[{{Kraichnan}(1968)}]{kraichnan68}
{Kraichnan}, R.~H. 1968, Physics of Fluids, 11, 945, \dodoi{10.1063/1.1692063}

\bibitem[{{Kulsrud} \& {Anderson}(1992)}]{kulsrud92}
{Kulsrud}, R.~M., \& {Anderson}, S.~W. 1992, \apj, 396, 606,
  \dodoi{10.1086/171743}

\bibitem[{{Larson} {et~al.}(2023){Larson}, {Finkelstein}, {Kocevski},
  {Hutchison}, {Trump}, {Haro}, {Bromm}, {Cleri}, {Dickinson}, {Fujimoto},
  {Kartaltepe}, {Koekemoer}, {Papovich}, {Pirzkal}, {Tacchella}, {Zavala},
  {Bagley}, {Behroozi}, {Champagne}, {Cole}, {Jung}, {Morales}, {Yang},
  {Zhang}, {Zitrin}, {Amor{\'\i}n}, {Burgarella}, {Casey}, {Ch{\'a}vez Ortiz},
  {Cox}, {Chworowsky}, {Fontana}, {Gawiser}, {Grazian}, {Grogin}, {Harish},
  {Hathi}, {Hirschmann}, {Holwerda}, {Juneau}, {Leung}, {Lucas}, {McGrath},
  {P{\'e}rez-Gonz{\'a}lez}, {Rigby}, {Seill{\'e}}, {Simons}, {de La Vega},
  {Weiner}, {Wilkins}, {Yung}, \& {Ceers Team}}]{larson23}
{Larson}, R.~L., {Finkelstein}, S.~L., {Kocevski}, D.~D., {et~al.} 2023, \apjl,
  953, L29, \dodoi{10.3847/2041-8213/ace619}

\bibitem[{{Latif} \& {Schleicher}(2023)}]{latif23b}
{Latif}, M.~A., \& {Schleicher}, D. R.~G. 2023, \apjl, 952, L9,
  \dodoi{10.3847/2041-8213/ace34f}

\bibitem[{{Latif} {et~al.}(2023){Latif}, {Schleicher}, \&
  {Khochfar}}]{latif23a}
{Latif}, M.~A., {Schleicher}, D. R.~G., \& {Khochfar}, S. 2023, \apj, 945, 137,
  \dodoi{10.3847/1538-4357/acbcc2}

\bibitem[{{Lesur}(2021)}]{lesur21}
{Lesur}, G. R.~J. 2021, \aap, 650, A35, \dodoi{10.1051/0004-6361/202040109}

\bibitem[{{Lodato}(2007)}]{lodato07}
{Lodato}, G. 2007, Nuovo Cimento Rivista Serie, 30, 293,
  \dodoi{10.1393/ncr/i2007-10022-x}

\bibitem[{{Luo} {et~al.}(2018){Luo}, {Ardaneh}, {Shlosman}, {Nagamine}, {Wise},
  \& {Begelman}}]{luo18}
{Luo}, Y., {Ardaneh}, K., {Shlosman}, I., {et~al.} 2018, \mnras, 476, 3523,
  \dodoi{10.1093/mnras/sty362}

\bibitem[{{Luo} {et~al.}(2023){Luo}, {Shlosman}, \& {Nagamine}}]{luo23}
{Luo}, Y., {Shlosman}, I., \& {Nagamine}, K. 2023, \apj, 955, 99,
  \dodoi{10.3847/1538-4357/acefb9}

\bibitem[{{Machida} {et~al.}(2006){Machida}, {Nakamura}, \&
  {Matsumoto}}]{machida06}
{Machida}, M., {Nakamura}, K.~E., \& {Matsumoto}, R. 2006, \pasj, 58, 193,
  \dodoi{10.1093/pasj/58.1.193}

\bibitem[{{Maiolino} {et~al.}(2024){Maiolino}, {Scholtz}, {Witstok},
  {Carniani}, {D'Eugenio}, {de Graaff}, {{\"U}bler}, {Tacchella},
  {Curtis-Lake}, {Arribas}, {Bunker}, {Charlot}, {Chevallard}, {Curti},
  {Looser}, {Maseda}, {Rawle}, {Rodr{\'\i}guez del Pino}, {Willott}, {Egami},
  {Eisenstein}, {Hainline}, {Robertson}, {Williams}, {Willmer}, {Baker},
  {Boyett}, {DeCoursey}, {Fabian}, {Helton}, {Ji}, {Jones}, {Kumari},
  {Laporte}, {Nelson}, {Perna}, {Sandles}, {Shivaei}, \& {Sun}}]{maiolino23}
{Maiolino}, R., {Scholtz}, J., {Witstok}, J., {et~al.} 2024, \nat, 627, 59,
  \dodoi{10.1038/s41586-024-07052-5}

\bibitem[{{Matsumoto} {et~al.}(1988){Matsumoto}, {Horiuchi}, {Shibata}, \&
  {Hanawa}}]{matsumoto88}
{Matsumoto}, R., {Horiuchi}, T., {Shibata}, K., \& {Hanawa}, T. 1988, \pasj,
  40, 171

\bibitem[{{Mishra} {et~al.}(2020){Mishra}, {Begelman}, {Armitage}, \&
  {Simon}}]{mishra20}
{Mishra}, B., {Begelman}, M.~C., {Armitage}, P.~J., \& {Simon}, J.~B. 2020,
  \mnras, 492, 1855, \dodoi{10.1093/mnras/stz3572}

\bibitem[{{Noble} {et~al.}(2010){Noble}, {Krolik}, \& {Hawley}}]{noble2010}
{Noble}, S.~C., {Krolik}, J.~H., \& {Hawley}, J.~F. 2010, ApJ, 711, 959,
  \dodoi{10.1088/0004-637X/711/2/959}

\bibitem[{{Norman} \& {Bryan}(1999)}]{norman99}
{Norman}, M.~L., \& {Bryan}, G.~L. 1999, in Astrophysics and Space Science
  Library, Vol. 240, Numerical Astrophysics, ed. S.~M. {Miyama}, K.~{Tomisaka},
  \& T.~{Hanawa}, 19, \dodoi{10.1007/978-94-011-4780-4_3}

\bibitem[{{Novikov} \& {Thorne}(1973)}]{novikov73}
{Novikov}, I.~D., \& {Thorne}, K.~S. 1973, in Black Holes (Les Astres Occlus),
  ed. C.~{Dewitt} \& B.~S. {Dewitt}, 343--450

\bibitem[{{Pariev}(2001)}]{pariev01}
{Pariev}, V.~I. 2001, PhD thesis, University of Arizona

\bibitem[{{Parker}(1955)}]{parker55}
{Parker}, E.~N. 1955, \apj, 122, 293, \dodoi{10.1086/146087}

\bibitem[{{Parker}(1966)}]{parker66}
---. 1966, \apj, 145, 811, \dodoi{10.1086/148828}

\bibitem[{{Pessah} \& {Psaltis}(2005)}]{pessah05}
{Pessah}, M.~E., \& {Psaltis}, D. 2005, \apj, 628, 879, \dodoi{10.1086/430940}

\bibitem[{{Planck Collaboration} {et~al.}(2016){Planck Collaboration}, {Ade},
  {Aghanim}, {Arnaud}, {Ashdown}, {Aumont}, {Baccigalupi}, {Banday},
  {Barreiro}, {Bartlett}, \& et~al.}]{planck-collaboration16}
{Planck Collaboration}, {Ade}, P.~A.~R., {Aghanim}, N., {et~al.} 2016, \aap,
  594, A13, \dodoi{10.1051/0004-6361/201525830}

\bibitem[{{Prole} {et~al.}(2024){Prole}, {Regan}, {Glover}, {Klessen},
  {Priestley}, \& {Clark}}]{prole24}
{Prole}, L.~R., {Regan}, J.~A., {Glover}, S. C.~O., {et~al.} 2024, \aap, 685,
  A31, \dodoi{10.1051/0004-6361/202348903}

\bibitem[{{Pudritz} \& {Silk}(1989)}]{pudritz89}
{Pudritz}, R.~E., \& {Silk}, J. 1989, \apj, 342, 650, \dodoi{10.1086/167625}

\bibitem[{{Ruzmaikin} {et~al.}(1988){Ruzmaikin}, {Sokolov}, \&
  {Shukurov}}]{ruzmaikin88}
{Ruzmaikin}, A.~A., {Sokolov}, D.~D., \& {Shukurov}, A.~M. 1988, {Magnetic
  Fields of Galaxies}, Vol. 133, \dodoi{10.1007/978-94-009-2835-0}

\bibitem[{{Ruzmaikin} {et~al.}(1979){Ruzmaikin}, {Turchaninov}, {Zeldovich}, \&
  {Sokoloff}}]{ruzmaikin79}
{Ruzmaikin}, A.~A., {Turchaninov}, V.~I., {Zeldovich}, I.~B., \& {Sokoloff},
  D.~D. 1979, \apss, 66, 369, \dodoi{10.1007/BF00650011}

\bibitem[{{Shakura} \& {Sunyaev}(1973)}]{shakura73}
{Shakura}, N.~I., \& {Sunyaev}, R.~A. 1973, A\&A, 24, 337

\bibitem[{{Sharda} {et~al.}(2021){Sharda}, {Federrath}, {Krumholz}, \&
  {Schleicher}}]{sharda21}
{Sharda}, P., {Federrath}, C., {Krumholz}, M.~R., \& {Schleicher}, D. R.~G.
  2021, \mnras, 503, 2014, \dodoi{10.1093/mnras/stab531}

\bibitem[{{Shlosman} \& {Begelman}(1987)}]{shlosman87}
{Shlosman}, I., \& {Begelman}, M.~C. 1987, \nat, 329, 810,
  \dodoi{10.1038/329810a0}

\bibitem[{{Shlosman} \& {Begelman}(1989)}]{shlosman89a}
---. 1989, \apj, 341, 685, \dodoi{10.1086/167526}

\bibitem[{{Shlosman} {et~al.}(1990){Shlosman}, {Begelman}, \&
  {Frank}}]{shlosman90}
{Shlosman}, I., {Begelman}, M.~C., \& {Frank}, J. 1990, \nat, 345, 679,
  \dodoi{10.1038/345679a0}

\bibitem[{{Shlosman} {et~al.}(2016){Shlosman}, {Choi}, {Begelman}, \&
  {Nagamine}}]{shlosman16}
{Shlosman}, I., {Choi}, J.-H., {Begelman}, M.~C., \& {Nagamine}, K. 2016,
  \mnras, 456, 500, \dodoi{10.1093/mnras/stv2700}

\bibitem[{{Shlosman} {et~al.}(1989){Shlosman}, {Frank}, \&
  {Begelman}}]{shlosman89}
{Shlosman}, I., {Frank}, J., \& {Begelman}, M.~C. 1989, \nat, 338, 45,
  \dodoi{10.1038/338045a0}

\bibitem[{{Silk} \& {Langer}(2006)}]{silk06}
{Silk}, J., \& {Langer}, M. 2006, \mnras, 371, 444,
  \dodoi{10.1111/j.1365-2966.2006.10689.x}

\bibitem[{{Smith} {et~al.}(2017){Smith}, {Bryan}, {Glover}, {Goldbaum}, {Turk},
  {Regan}, {Wise}, {Schive}, {Abel}, {Emerick}, {O'Shea}, {Anninos}, {Hummels},
  \& {Khochfar}}]{smith17}
{Smith}, B.~D., {Bryan}, G.~L., {Glover}, S.~C.~O., {et~al.} 2017, \mnras, 466,
  2217, \dodoi{10.1093/mnras/stw3291}

\bibitem[{{Steinwandel} {et~al.}(2019){Steinwandel}, {Beck}, {Arth}, {Dolag},
  {Moster}, \& {Nielaba}}]{steinwandel19}
{Steinwandel}, U.~P., {Beck}, M.~C., {Arth}, A., {et~al.} 2019, \mnras, 483,
  1008, \dodoi{10.1093/mnras/sty3083}

\bibitem[{{Subramanian}(1998)}]{subramanian98}
{Subramanian}, K. 1998, \mnras, 294, 718,
  \dodoi{10.1046/j.1365-8711.1998.01284.x10.1111/j.1365-8711.1998.01284.x}

\bibitem[{{Sur} {et~al.}(2008){Sur}, {Brandenburg}, \& {Subramanian}}]{sur08}
{Sur}, S., {Brandenburg}, A., \& {Subramanian}, K. 2008, \mnras, 385, L15,
  \dodoi{10.1111/j.1745-3933.2008.00423.x}

\bibitem[{{Suzuki} \& {Inutsuka}(2014)}]{suzuki14}
{Suzuki}, T.~K., \& {Inutsuka}, S.-i. 2014, \apj, 784, 121,
  \dodoi{10.1088/0004-637X/784/2/121}

\bibitem[{{Tan} \& {Blackman}(2004)}]{tan04}
{Tan}, J.~C., \& {Blackman}, E.~G. 2004, \apj, 603, 401, \dodoi{10.1086/381668}

\bibitem[{{Toomre}(1964)}]{toomre64}
{Toomre}, A. 1964, \apj, 139, 1217, \dodoi{10.1086/147861}

\bibitem[{{Truelove} {et~al.}(1997){Truelove}, {Klein}, {McKee}, {Holliman},
  {Howell}, \& {Greenough}}]{truelove97}
{Truelove}, J.~K., {Klein}, R.~I., {McKee}, C.~F., {et~al.} 1997, \apj, 489,
  L179, \dodoi{10.1086/310975}

\bibitem[{{Turk} {et~al.}(2009){Turk}, {Abel}, \& {O'Shea}}]{turk09}
{Turk}, M.~J., {Abel}, T., \& {O'Shea}, B. 2009, Science, 325, 601,
  \dodoi{10.1126/science.1173540}

\bibitem[{{Turk} {et~al.}(2011){Turk}, {Smith}, {Oishi}, {Skory}, {Skillman},
  {Abel}, \& {Norman}}]{turk11}
{Turk}, M.~J., {Smith}, B.~D., {Oishi}, J.~S., {et~al.} 2011, The Astrophysical
  Journal Supplement Series, 192, 9, \dodoi{10.1088/0067-0049/192/1/9}

\bibitem[{{van Leer}(1979)}]{vanleer79}
{van Leer}, B. 1979, JCoPh, 32, 101, \dodoi{10.1016/0021-9991(79)90145-1}

\bibitem[{{Velihov}(1959)}]{velihov59}
{Velihov}, E. 1959, Soviet.Phys.JETP, 36, 1398

\bibitem[{{Venemans} {et~al.}(2017){Venemans}, {Walter}, {Decarli},
  {Ba{\~n}ados}, {Carilli}, {Winters}, {Schuster}, {da Cunha}, {Fan}, {Farina},
  {Mazzucchelli}, {Rix}, \& {Weiss}}]{venemans17}
{Venemans}, B.~P., {Walter}, F., {Decarli}, R., {et~al.} 2017, \apjl, 851, L8,
  \dodoi{10.3847/2041-8213/aa943a}

\bibitem[{{Woods} {et~al.}(2021){Woods}, {Patrick}, {Elford}, {Whalen}, \&
  {Heger}}]{woods21}
{Woods}, T.~E., {Patrick}, S., {Elford}, J.~S., {Whalen}, D.~J., \& {Heger}, A.
  2021, ApJ, 915, 110, \dodoi{10.3847/1538-4357/abfaf9}

\bibitem[{{Woods} {et~al.}(2024){Woods}, {Patrick}, {Whalen}, \&
  {Heger}}]{woods24}
{Woods}, T.~E., {Patrick}, S., {Whalen}, D.~J., \& {Heger}, A. 2024, \apj, 960,
  59, \dodoi{10.3847/1538-4357/ad054a}

\bibitem[{{Wu} {et~al.}(2015){Wu}, {Wang}, {Fan}, {Yi}, {Zuo}, {Bian}, {Jiang},
  {McGreer}, {Wang}, {Yang}, {Yang}, {Thompson}, \& {Beletsky}}]{wu15}
{Wu}, X.-B., {Wang}, F., {Fan}, X., {et~al.} 2015, \nat, 518, 512,
  \dodoi{10.1038/nature14241}

\bibitem[{{Zeldovich}(1983)}]{zeldovich83}
{Zeldovich}, Y.~B. 1983, {Magnetic fields in astrophysics}

\bibitem[{{Zhu} \& {Stone}(2018)}]{zhu18}
{Zhu}, Z., \& {Stone}, J.~M. 2018, \apj, 857, 34,
  \dodoi{10.3847/1538-4357/aaafc9}

\bibitem[{{Ziegler}(2001)}]{ziegler01}
{Ziegler}, U. 2001, \aap, 367, 170, \dodoi{10.1051/0004-6361:20000404}

\end{thebibliography}
\bibliographystyle{aasjournal} 


\end{document}